\documentclass{jfm}
\usepackage{graphicx}
\usepackage{natbib}
\usepackage{url}

\ifCUPmtlplainloaded \else
  \checkfont{eurm10}
  \iffontfound
    \IfFileExists{upmath.sty}
      {\typeout{^^JFound AMS Euler Roman fonts on the system,
                   using the 'upmath' package.^^J}%
       \usepackage{upmath}}
      {\typeout{^^JFound AMS Euler Roman fonts on the system, but you
                   dont seem to have the}%
       \typeout{'upmath' package installed. JFM.cls can take advantage
                 of these fonts,^^Jif you use 'upmath' package.^^J}%
      }
  \else
  \fi
\fi

\ifCUPmtlplainloaded \else
  \checkfont{msam10}
  \iffontfound
    \IfFileExists{amssymb.sty}
      {\typeout{^^JFound AMS Symbol fonts on the system, using the
                'amssymb' package.^^J}%
       \usepackage{amssymb}%
         \let\leq=\leqslant
         
      }{}
  \fi
\fi

\ifCUPmtlplainloaded \else
  \IfFileExists{amsbsy.sty}
    {\typeout{^^JFound the 'amsbsy' package on the system, using it.^^J}%
     \usepackage{amsbsy}}
    {\providecommand\boldsymbol[1]{\mbox{\boldmath $##1$}}}
\fi

\usepackage{amssymb, epsfig,amssymb, latexsym}
\usepackage{amsfonts,psfrag,amsmath,bbm,color}
\usepackage{lscape}

\newcommand{\md}{\mathbb{D}}

\renewcommand{\phi}{\varphi}

\newcommand{\obu}{\overline{\bf u}}
\newcommand{\tbu}{\widetilde{\bf u}}
\newcommand{\tobu}{\widetilde{\overline{\bf u}}}

\newcommand{\bu}{{\bf u}}
\newcommand{\bk}{{\bf k}}

\newcommand{\op}{\overline{p}}

\newcommand{\bx}{{\bf x}}
\newcommand{\bX}{{\bf X}}

\newcommand{\bT}{{\bf T}}
\newcommand{\bL}{{\bf L}}

\def\PP{{{\rm l}\kern - .15em {\rm P} }}
\def\PN2{{\PP_{N}-\PP_{N-2}}}

\newcommand{\D}{\mathbbm{D}}

\newcommand{\btau}{\boldsymbol{\tau}}
\newcommand{\bphi}{\boldsymbol{\phi}}
\newcommand{\bvarphi}{\boldsymbol{\varphi}}

\newcommand{\ba}{{\bf a}}

\title[POD Closure Models for Turbulent Flows]{Proper Orthogonal Decomposition \\ 
	Closure Models For 
	Turbulent Flows: \\ 
	A Numerical Comparison}

\author[Zhu Wang, Imran Akhtar, Jeff Borggaard, and Traian Iliescu]{Zhu Wang$^1$, Imran Akhtar$^2$, Jeff Borggaard$^1$, and Traian Iliescu$^1$}

\affiliation{$^1$Department of Mathematics, Virginia Tech, Blacksburg, VA 24061-0123, U.S.A.\\[\affilskip]
$^2$Department of Mechanical Engineering, NUST College of Electrical \& Mechanical Engineering, National University of Sciences \& Technology, Islamabad, Pakistan.}

\pubyear{1996}
\volume{538}
\pagerange{119--126}
\date{$\today$ and in revised form ??}

\begin{document}

\maketitle

\begin{abstract}
This paper puts forth two new closure models for the proper orthogonal decomposition
reduced-order modeling 
of structurally dominated  turbulent flows: the dynamic subgrid-scale model and the
variational multiscale model.
These models, which are considered state-of-the-art in large eddy simulation,
together with the mixing length and the Smagorinsky closure models, are tested in the
numerical simulation 
of a 3D turbulent
flow around a circular cylinder at $Re = 1,000$.                                      
Two criteria are used in judging the performance of the proper orthogonal decomposition 
reduced-order 
models: the kinetic energy spectrum and the time evolution of the POD coefficients.
All the numerical results are benchmarked against a direct numerical simulation.
Based on these numerical results, we conclude that the dynamic subgrid-scale  and 
the variational multiscale models perform best.
\end{abstract}


\section{Introduction}
\label{s_introduction}

\textit{Reduced-order models (ROMs)} of structurally dominated turbulent flows are central to many 
applications in science and engineering, such as 
fluid flow control 
\cite[see for example][]{ito1998reduced,graham1999ocv2,cohen2003fcc,bergmann2005optimal,lehmann2005wsp,hoepffner2006ccd,bagheri2009input,barbagallo2009closed,ahuja2010feedback,akhtar2010model}
and data assimilation of atmospheric and oceanic flows 
\cite[]{LZWN07,daescu2008dual,fang2009pod}.
Both computational efficiency and physical accuracy are needed for the 
success of these ROMs in practical applications.
Striking a balance between efficiency and accuracy in ROMs of turbulent flows
is, of course, challenging.
Indeed, it is clear that the fewer the modes retained in the ROM, the more efficient 
the ROM is.
Preserving the physical accuracy of the resulting ROM, however, becomes challenging,
since the modes that are not retained in the ROM representation of the underlying 
turbulent flow need to be modeled.
The quest of balancing the computational efficiency and physical accuracy represents
one of the main challenges in ROMs for turbulent flows. 

One of the most successful ROM strategies for structurally dominated turbulent flows 
has been the \textit{Proper Orthogonal Decomposition (POD)} 
\cite[see for example][]{HLB96,Sir87abc}. 
POD starts with data from an accurate numerical simulation (or physical experiment), 
extracts the most energetic modes in the system, 
and utilizes a Galerkin procedure that yields a ROM of the underlying turbulent flow.
The first {\em proper orthogonal decomposition reduced-order model (POD-ROM)} 
for the turbulent boundary layer was proposed in \cite{AHLS88}.
This model truncated the POD basis and used an eddy viscosity-based
approximation to model the effect of the discarded POD modes on the
POD modes kept in the model.
This POD-ROM yielded good qualitative 
results, considering the coarseness of the approximation.
The criterion used to assess the accuracy of the model was the 
intermittency of bursting events in the turbulent boundary layer.
This POD-ROM was further investigated numerically in two subsequent 
papers~\cite[]{PL98,Pod01}.
The model reproduced the qualitative physics of the turbulent boundary 
layer well.
Furthermore, by adding new POD modes to the model, the accuracy of
the model was increased. 

Despite their initial success, POD-ROMs have generally been limited 
to laminar flows and relatively few reports on closure 
modeling strategies for turbulent flows have appeared in
the literature
 \cite[]{AHLS88,PL98,Pod01,RF94,rempfer1996investigations,cazemier1998proper,ma2002low,sirisup2004spectral,buffoni2006low,noack2002low,NAMTT03,noack2005need,noack2008finite,ullmann2010pod,hay2009local,hay2010reduced}.
This is in stark contrast to the amount of work done in traditional
turbulence modeling, such as {\em large eddy simulation (LES)}, 
where literally hundreds of closure 
models have been proposed and investigated \cite[see for example][]{Sag06} 
over the same time period.
This disparity in closure modeling between POD reduced-order modeling 
and classical 
turbulence modeling seems even more dramatic considering that the
concept of an energy cascade, which is a fundamental modeling principle in
LES, is also valid in a POD setting.
Indeed, the validity of the extension of the energy cascade concept 
to the POD setting was investigated numerically in \cite{CSB03}.
The authors have investigated the energy transfer among POD modes
in a non-homogeneous computational setting.
By monitoring the triad interactions due to the nonlinear
term in the Navier-Stokes equations, they have concluded that the
transfer of energy among the POD modes is similar to the transfer
of energy among Fourier modes.
Specifically, they found that there is a net forward energy 
transfer from low index POD modes to higher index POD modes
and that this transfer of energy is local in nature
(that is, energy is mainly transferred among POD modes whose indices
are close to one another).
This study \cite[see also][]{noack2002low} clearly suggests 
that LES ideas based on the energy cascade concept could also be used 
in devising POD-ROMs.

One of the main reasons for the scarcity of closure models for POD-ROMs
of turbulent flows is the impractical cost of standard LES closure models
employed in a POD-ROM setting.
Indeed, most of the computational cost of a POD-ROM lies in assembling
the vectors, matrices and tensors of the ROM.
This, however, is hardly a problem for POD-ROM, since the vectors, matrices
and tensors are assembled only {\em once}, at the beginning of the POD-ROM
simulation, and reused at every time step.
Standard (nonlinear) LES closure models, however, 
introduce new vectors and matrices that need to be recomputed at 
every time step.
Thus, a straightforward numerical discretization of such closure models
would come at a huge computational cost, rendering the resulting POD-ROMs
impractical.

In the past few years, a number of strategies have been introduced
to treat nonlinear terms in POD-ROMs.  
These include interpolatory methods such as 
the empirical interpolation method 
\cite[]{barrault2004eim,chaturantabut2010nonlinear,galbally2010non}, 
the closely related group finite element approach \cite[]{dickinson2010nmr} and 
a novel two-level discretization method \cite[]{wang2011two}.  
The latter approach is best suited for this study since it does not
constrain the nonlinear term to lie within a predefined set.
This approach is based on a two-level discretization of the vectors, 
matrices and tenstors of the POD-ROM, in which all the terms are computed 
on the fine grid, except for the nonlinear closure model terms, which are
computed on a coarser grid.
In \cite{wang2011two}, numerical simulations of a turbulent flow past a 
3D cylinder at $Re = 1,000$ with a standard LES closure model \cite[]{Sma63} have shown that the new two-level discretization is both
computationally efficient and physically accurate.
Indeed, the new two-level algorithm decreased by more than an order of
magnitude the CPU time of the standard one-level algorithm, without 
compromising the physical accuracy.

In this report, we use the two-level algorithm proposed in \cite{wang2011two}
to discretize two new POD-ROMs, inspired from state-of-the-art LES closure modeling
strategies: the {\em dynamic subgrid-scale (DS)
model} \cite[]{GPMC91,MLC96,porte2000scale} and the {\em variational multiscale (VMS)} model 
\cite[]{HMJ00}.
We also consider the standard mixing-length closure model proposed in 
\cite{AHLS88} and the Smagorinsky model proposed in \cite{wang2011two} \cite[see also][]{noack2002low,ullmann2010pod}, 
both being standard LES closure models.
All four POD-ROMs are tested in the numerical simulation of a 3D turbulent
flow around a circular cylinder at $Re = 1,000$.                                      
Two criteria are used in judging the performance of the POD-ROMs: the kinetic energy spectrum and the time evolution of the POD coefficients.
All the numerical results are benchmarked against a direct numerical simulation.

The rest of the paper is organized as follows:
The general methodology used in the development of POD-ROMs
is presented in \S\,\ref{s_pod_rom}.
The four POD closure models are described in \S\,\ref{s_closure_models}
and are investigated numerically in \S\,\ref{s_numerical_results}.
Finally, conclusions and several research directions currently pursued by 
our group are provided in \S\,\ref{s_conclusions}.

\section{POD Reduced-Order Modeling}
\label{s_pod_rom}

We now present the general approach used in the development of POD-ROMs.
We start by briefly describing the POD methodology. 
For more details, the reader 
is referred to \cite{Sir87abc,HLB96}.
To this end, we consider the numerical solution of the incompressible 
{\it Navier-Stokes equations (NSE)}:
\begin{equation}
\label{NSE}
\left. \begin{array}{r}\textbf{u}_{t} - \mbox{Re}^{-1} \Delta \textbf{u} + (\textbf{u} \cdot \nabla) \textbf{u} + \nabla p = 0\\
\nabla \cdot \textbf{u} = 0 , \end{array} \right\}
\end{equation}
where $\textbf{u}$ is the velocity,
$p$ the pressure and
$\mbox{Re}$ the Reynolds number.
The POD basis is generated by post-processing typical
data from the numerical simulation of \eqref{NSE}.
If ${\cal Y} = \left\{ \textbf{y}(\cdot,t)\in {\cal H}\ |\ t\in(0,T) \right\}$
(with ${\cal H}$ a Hilbert space) represents a simulation of the NSE,
then the first POD basis vector is the function that maximizes the time-averaged 
projection of ${\cal Y}$ onto itself,
\begin{eqnarray}
\label{eq:phi1}
  {\boldsymbol \phi}_1 = \max_{{\boldsymbol \phi}\in{\cal H}, \| {\boldsymbol \phi} \|_{\cal H} = 1}
  \frac{1}{T} \int_0^T \left|\left\langle \textbf{y}(\cdot,t), {\boldsymbol \phi}(\cdot) \right\rangle_{\cal H}
  \right|^2 \ dt.
\end{eqnarray}
Subsequent vectors, ${\boldsymbol \phi}_k$, are determined by seeking the above maximum in the orthogonal
complement to
\begin{eqnarray}
\label{eq:Phik-1}
  {\bf X}^{k-1} = \mbox{span}\{{\boldsymbol \phi}_1, \ldots, {\boldsymbol \phi}_{k-1} \}, \quad 2 \leq k \leq N, \quad \mbox{in} \quad {\cal H},
\end{eqnarray}
where $N$ is the rank of ${\cal Y}$. 
If we choose
${\cal H} = {\cal L}_2$ and ${\cal Y}$ represents a single simulation, the POD basis functions
satisfy the Fredholm integral equation
\begin{eqnarray}
\label{eq:fredholm}
  \int_\Omega \textbf{R}({\bf x},{\bf x}^\prime) {\boldsymbol \phi}_i({\bf x}^\prime) \ d{\bf x}^\prime = 
  \lambda_i {\boldsymbol \phi}_i({\bf x}) ,
\end{eqnarray}
\noindent where
\begin{eqnarray}
\label{eq:kernel}
 \textbf{R}({\bf x},{\bf x}^\prime) = \frac{1}{T} \int_0^T \textbf{y}({\bf x},t) \textbf{y}^*({\bf x}^\prime,t)\ dt
\end{eqnarray}
\noindent is the spatial autocorrelation kernel.  There are natural extensions of this
definition that accommodate multiple simulations.  In practice, either the time average of
each simulation or the steady state solution is removed, so that ${\cal Y}$ contains 
fluctuation from the mean (or a centering trajectory), e.g.,
$\textbf{y}({\bf x},t) = {\bf u}({\bf x},t) - {\bf U}({\bf x})$ \cite[]{HLB96}.  
Note that each POD basis vector ${\boldsymbol \phi}_k$ represents a weighted time average of the data ${\cal Y}$.  Thus, these basis vectors
preserve linear properties (such as the divergence-free property). 

A POD basis enables a reduced representation of the simulated data, and thus can be viewed as a 
compression algorithm. 
Utilizing the POD basis to obtain efficient approximations to (\ref{NSE}) is achieved using the 
POD basis in a Galerkin approximation, and employing the fact that the POD basis vectors are 
mutually orthogonal.  
A POD-ROM of the flow is constructed from the POD basis by writing
\begin{equation}
\label{eq:yr}
  {\bf u}({\bf x},t) \approx {\bf u}_r({\bf x},t) \equiv {\bf U}({\bf x}) + \sum_{j=1}^r a_j(t) \boldsymbol\phi_j({\bf x}) ,
\end{equation}
where ${\bf U}({\bf x})$ is the centering trajectory, $\{ \boldsymbol \phi_{j}\}_{j=1}^{r}$ are the first 
$r$ POD basis vectors, and $\{a_{j}(t)\}_{j=1}^{r}$ are the sought time-varying coefficients that 
represent the POD-Galerkin trajectories.
We now replace the velocity $\bu$ with $\bu_{r}$ in the NSE
\eqref{NSE}, and then project the resulting equations onto the subspace $\bX^r$.
Using the boundary conditions and the fact that all modes are solenoidal,
one obtains the {\em POD Galerkin reduced-order model (POD-G-ROM)}:
\begin{eqnarray}
\left( \frac{\partial \bu_{r}}{\partial t} , \bphi \right)
+ \left( (\bu_{r} \cdot \nabla) \bu_{r} , \bphi \right)
+ \left( \frac{2}{Re} \, \D(\bu_{r}) , \nabla \bphi \right)
= 0 \quad
\forall \, \bphi \in \bX^r,
\label{pod_g}
\end{eqnarray}
where
$\D(\bu_{r}) := (\nabla \bu^r + (\nabla \bu^r)^T) / 2$ is the deformation tensor 
of $\bu_{r}$.
We note that, since the computational domain that we consider is large enough,
the pressure terms in \eqref{pod_g} can be neglected \cite[for details, see][]{noack2005need,akhtar2009stability}. 
The POD-G-ROM \eqref{pod_g} yields the following autonomous dynamical system for the
vector of time coefficients, ${\bf a}(t)$:
\begin{equation}
\label{eq:a}
  \dot{\bf a} = {\bf b} + {\bf A} {\bf a}   + {\bf a}^T {\bf B} {\bf a} ,
\end{equation}
\noindent 
where $\bf{b}$, $\bf{A}$, and $\bf{B}$ correspond to the constant, linear, and quadratic terms in
the numerical discretization of the NSE \eqref{NSE}, respectively. The initial conditions are obtained by projection:
\begin{equation}
\label{eq:initial}
a_j(0)=\langle {\boldsymbol \phi}_j, {\bu}(\cdot, 0)-{\bf U}(\cdot)\rangle_{\cal H}, \quad j=1, \ldots, r.
\end{equation}
The finite dimensional system \eqref{eq:a} can be written componentwise as follows:
For all $k = 1, \ldots, r$,
\begin{eqnarray}
\label{podg_k}
\dot{a}_k(t) 
=  b_k
+ \sum_{m=1}^{r} A_{km}a_m(t) + \sum_{m=1}^r \sum_{n=1}^r B_{kmn}a_n(t)a_m(t),
\end{eqnarray}
where
\begin{eqnarray}
&& \hspace*{-1.3cm} b_k 
= -\left( \boldsymbol\phi_k, {\bf U} \cdot \nabla {\bf U} \right) 
- \frac{2}{\mbox{Re}} \left( \nabla \boldsymbol\phi_k, \frac{\nabla {\bf U} +\nabla {\bf U}^{T}}{2} \right), \label{podg_bk}\\
&& \hspace*{-1.3cm} A_{km} 
= -(\boldsymbol\phi_k, {\bf U} \cdot \nabla \boldsymbol\phi_m) 
- (\boldsymbol\phi_k, \boldsymbol\phi_m \cdot \nabla {{\bf U}}) 
- \frac{2}{\mbox{Re}} \left( \nabla \boldsymbol\phi_k, \frac{\nabla \boldsymbol\phi_m 
+ \nabla {\boldsymbol\phi_m}^{T}}{2} \right), \label{podg_Ak}\\
&& \hspace*{-1.3cm} B_{kmn}= -(\boldsymbol\phi_k, \boldsymbol\phi_m \cdot \nabla \boldsymbol\phi_n). \label{podg_Bk}
\end{eqnarray}

\section{POD Closure Models}
\label{s_closure_models}

In this section, we present the four POD closure models investigated numerically in 
\S\,\ref{s_numerical_results}.
To this end, we start by describing the filtering operation utilized and the spatial
lengthscale $\delta$ used in the POD closure models.
Both are needed in order to define meaningful LES-inspired POD closure models.

\subsection{POD Filter}
\label{s_pod_filtering}

In LES, the filter is the central tool used to obtain simplified mathematical 
models that are computationally tractable.
The filtering operation is effected by convolution of flow variables with a rapidly decaying 
{\em spatial} filter $g_{\delta}$, where $\delta$ is the radius of the spatial filter.
In POD, however, there is no explicit spatial filter used.
Thus, in order to develop LES-type POD closure models, a POD filter needs
to be introduced.
Given the hierarchical nature of the POD basis, a natural such filter appears to
be the Galerkin projection.
For all ${\bf u} \in \bX$, the Galerkin projection $\obu \in \bX^r$ is the solution of the following equation:
\begin{eqnarray}
(\bu - \obu , \bphi) = 0
\qquad \forall \, \bphi \in \bX^r.
\label{galerkin_projection}
\end{eqnarray}
The Galerkin projection defined in \eqref{galerkin_projection} will be the filter 
used in all POD closure models studied in this report.

\subsection{POD Lengthscale}
\label{ss_pod_lengthscale}

Next, we introduce the lengthscale $\delta$ used in the POD closure models.
We emphasize that this choice is one of the fundamental issues in making a 
connection with LES.
Indeed, we need such a lengthscale ($\delta$) in order to define 
dimensionally sound POD models of LES flavor. 

To derive the lengthscale $\delta$, we use dimensional analysis.
\cite{AHLS88} defined $l_{>}$, a dimensionally sound 
lengthscale for a turbulent pipe flow.
In fact, this lengthscale was only defined implicitly, through the turbulent eddy viscosity
$\nu_{T} := u_{>} \, l_{>}$.
Indeed, equation $(22)$ in \cite{AHLS88} reads
\begin{eqnarray}
\nu_{T} 
:= u_{>} \, l_{>}
= \frac{\int_{0}^{X_2} \langle u_{i >}  \, u_{i >} \rangle \, dx_2}
{\left( X_2 \, \int_{0}^{X_2} \langle u_{i >, j} \, u_{i >, j} \rangle \, dx_2 \right)^{1/2}} \, , 
\label{nu_T_AHLS88}
\end{eqnarray}
where repeated indices denote summation, 
the subscript $_{>}$ denotes unresolved POD modes,
\begin{eqnarray}
\langle f \rangle = \frac{1}{L_1 \, L_3} \, \int_{0}^{L_1} \int_{0}^{L_3} 
f(\bx, t) \, dx_1 \, dx_3 
\label{spatial_average}
\end{eqnarray}
denotes the spatial average of $f$ in the homogeneous directions (here $x_1$ and $x_3$), 
and $L_1, L_3$ and $X_2$ are the streamwise, spanwise, and wall-normal dimensions of the 
computational domain, respectively.
Note that the authors only considered the wall region, not the entire pipe flow.
In \eqref{nu_T_AHLS88}, the following notation was used:
$\displaystyle 
u_{i >} 
= \sum_{j=r+1}^{N} a_j^i \, \varphi_j ,
\ 
u_{i >} \, u_{i >} 
= \sum_{i=1}^{3} u_{i >} \, u_{i >} , 
\ \text{and} \  
u_{i >, j}
= \frac{\partial u_{i >}}{\partial x_j} .
$
Note that a quick dimensional analysis shows that the quantity defined in \eqref{nu_T_AHLS88}
has the dimensions of a viscosity.
Indeed,
\begin{eqnarray}
[ \nu_{T} ] 
= \frac{\frac{m}{s} \, \frac{m}{s} \, m}
{\left[ m \, \left( \frac{1}{s} \, \frac{1}{s} \, m \right) \right]^{1/2}}
= \frac{\frac{m^3}{s^2}}{\frac{m}{s}}
= \frac{m^2}{s} \, .
\label{dimensional_analysis}
\end{eqnarray}

In Appendix B of \cite{AHLS88}, the authors have further simplified \eqref{nu_T_AHLS88}
and expressed $\nu_{T}$ in terms of the first neglected POD modes:
\begin{eqnarray}
\nu_{T} 
:= u_{>} \, l_{>}
= \frac{\sum_{(\bk, n)} \lambda_{\bk}^{(n)}}
{\left( X_2 \, L
_1 \, L_3 \, \sum_{(\bk, n)} \lambda_{\bk}^{(n)} \, 
\left( \int_{0}^{X_2}  D\Phi_{i_{\bk}}^{(n)} D\Phi_{i_{\bk}}^{(n) *} \, dx_2 
- k_1^2 - k_3^2 \right) \right)^{1/2}} \, , 
\label{nu_T_AHLS88_appendix_B}
\end{eqnarray}
where the triplets $(\bk, n)$ are the first neglected POD modes.

In equation $(9.90)$ in \cite{HLB96}, the authors define another dimensionally sound
turbulent viscosity
\begin{eqnarray}
\nu_{T} 
:= u_{>} \, l_{>}
= \frac{1}{X_2} \, \int_{0}^{X_2} \, 
  \frac{\langle u_{i >}  \, u_{i >} \rangle }
  {\langle u_{i >, j} \, u_{i >, j} \rangle^{1/2}} \, dx_2. 
\label{nu_T_HLB96}
\end{eqnarray}
A quick dimensional analysis shows that the quantity defined in \eqref{nu_T_HLB96}
also has the dimensions of a viscosity.

We can use the two definitions of $\nu_T$ in \eqref{nu_T_AHLS88} and \eqref{nu_T_HLB96}
to define a lengthscale $l_{>}$.
We obtain
\begin{eqnarray}
l_{>}
:= \frac{\int_{0}^{X_2} \langle u_{i >}  \, u_{i >} \rangle \, dx_2}
{X_2 \, \int_{0}^{X_2} \langle u_{i >, j} \, u_{i >, j} \rangle \, dx_2 }
\label{l_AHLS88}
\end{eqnarray}
and
\begin{eqnarray}
l_{>}
:= \left( \frac{1}{X_2} \, \int_{0}^{X_2} \, 
  \frac{\langle u_{i >}  \, u_{i >} \rangle }
  {\langle u_{i >, j} \, u_{i >, j} \rangle} \, dx_2 \right)^{1/2}\, , 
\label{l_HLB96}
\end{eqnarray}
respectively.

For our 3D flow past a cylinder, both \eqref{l_AHLS88} and \eqref{l_HLB96} are
valid candidates for the definition of the lengthscale $\delta$.
The only modification we need to make (due to our computational domain) is to replace 
the horizontal averaging by
spanwise averaging and take double integrals in the remaining directions.
Specifically, we have
\begin{eqnarray}
\delta
:= \left( \frac{\int_{0}^{L_1} \, \int_{0}^{L_2} \langle u_{i >}  \, u_{i >} \rangle \, dx_1 \, dx_2}
{\int_{0}^{L_1} \, \int_{0}^{L_2} \langle u_{i >, j} \, u_{i >, j} \rangle \, dx_1 \, dx_2 } \right)^{1/2}
\label{delta_AHLS88}
\end{eqnarray}
and
\begin{eqnarray}
\delta
:= \left( \frac{1}{L_1 \, L_2} \, \int_{0}^{L_1} \, \int_{0}^{L_2}  \, 
  \frac{\langle u_{i >}  \, u_{i >} \rangle }
  {\langle u_{i >, j} \, u_{i >, j} \rangle} \, dx_1 \, dx_2 \right)^{1/2}\, . 
\label{delta_HLB96}
\end{eqnarray}

\newpage
\subsection{POD Closure Models}
\label{ss_les}

We are now ready to present the four POD closure models that will be investigated numerically in 
\S\,\ref{s_numerical_results}.

The POD-G-ROM \eqref{pod_g} can be used 
for laminar flows.
For structurally dominated turbulent flows, however, the POD-G-ROM simply fails \cite[]{wang2011two}.
The reason is that the effect of the discarded POD modes $\{ \bphi_{r+1}, \ldots, \bphi_{N} \}$
needs to be included in the model.
For turbulent flows, the most natural way to tackle this POD {\em closure problem} is by using 
the {\em eddy viscosity (EV)} concept.
Indeed, most closure models used in turbulence modeling are based on this EV concept, which 
states that the role of the discarded modes is to extract energy from the system.
The concept of energy cascade, which is well established in a Fourier setting, has been recently
confirmed in a POD setting in the numerical investigations in \cite{CSB03}.
Thus, using LES inspired EV closure models in POD-ROM represents a natural step.

In this section, we propose two new POD closure models: the dynamic subgrid-scale model
and the variational multiscale model.
We emphasize that, although these models were announced in \cite{borggaard2008reduced},
this study represents their {\em first} careful derivation and thorough numerical investigation.
We also numerically test the mixing-length \cite[]{AHLS88} and Smagorinsky \cite[]{noack2002low,ullmann2010pod,wang2011two} POD closure models.

Since all four POD closure models are of EV type, we first present a general EV POD-ROM framework. 
Then, for each closure model, we specify the changes that need to be made to this general 
framework.
The general EV POD-ROM framework can be written as:
\begin{eqnarray}
\label{eq:a_closure}
\dot{\bf a} 
= \left( {\bf b} + \widetilde{\bf b}({\bf a}) \right)
+ \left( {\bf A} + \widetilde{\bf A}({\bf a}) \right) {\bf a} 
+ {\bf a}^T {\bf B} {\bf a} ,
\end{eqnarray}
which is just a slight modification of the POD-G-ROM \eqref{eq:a}.
The new terms in \eqref{eq:a_closure} (the vector $\widetilde{\bf b}({\bf a})$ and the matrix 
$\widetilde{\bf A}({\bf a})$) correspond to the numerical discretization of the POD closure model.
In componentwise form, equation \eqref{eq:a_closure} can be written as 
\begin{eqnarray}
\dot{a}_k(t) 
&=&  \left( b_k
+ \widetilde{b}_k({\bf a}) \right)
+ \sum_{m=1}^{r}  \left( A_{km} + \widetilde{A}_{km}({\bf a}) \right) a_m(t) \nonumber \\
&+& \sum_{m=1}^r \sum_{n=1}^r B_{kmn}a_n(t)a_m(t), \label{eq:a_closure_2}
\end{eqnarray}
where $b_k, A_{km}$, and $B_{kmn}$ are the same as those in equations \eqref{eq:a} and 
$\widetilde{b}_k({\bf a})$ and $\widetilde{A}_{km}({\bf a})$ depend on the specific closure
model used.

\subsubsection{The mixing-length POD reduced-order model (ML-POD-ROM)}
\label{sss_les_mixing_length}

The first POD closure model was the {\em mixing-length} model proposed in \cite{AHLS88}.
This closure model is of EV type and amounts to increasing the viscosity coefficient $\nu$ by
\begin{eqnarray}
\nu_{ML}
= \alpha \, \nu_T
= \alpha \, U_{ML} \, L_{ML},
\label{definition_mixing_length}
\end{eqnarray}
where $U_{ML}$ and $L_{ML}$ are characteristic velocity and length 
scales for the unresolved scales, and $\alpha$ is an $\mathcal{O}(1)$ nondimensional
parameter that characterizes the energy
being dissipated.
Using the EV ansatz in \eqref{definition_mixing_length}, the {\em mixing-length POD 
reduced-order model (ML-POD-ROM)} has the form in \eqref{eq:a_closure}, where
\begin{eqnarray}
\widetilde{b}_k({\bf a}) 
&=& -\nu_{ML}\left(\nabla {\boldsymbol\phi}_k,  \frac{\nabla{\bf U} 
+ \nabla {\bf U}^{T}}{2} \right), \label{ML_POD_ROM_1} \\
\widetilde{A}_{km}({\bf a}) 
&=& -\nu_{ML} \left(\nabla {\boldsymbol\phi}_k,  \frac{\nabla {\boldsymbol\phi}_m 
+ \nabla {{\boldsymbol\phi}_m}^{T}}{2} \right). \label{ML_POD_ROM_2}
\end{eqnarray}
The parameter $\alpha$ is expected to vary in a real turbulent
flow, and different values of $\alpha$ may result in different
dynamics of the flow \cite[]{AHLS88,HLB96,PL98,Pod01}.
There are also different ways to define $\nu_{T}$ in \eqref{eq:a_closure}: relation \eqref{nu_T_AHLS88} 
was used in \cite{AHLS88}, whereas relation \eqref{nu_T_HLB96} was used in \cite{HLB96}.
We also mention that several other authors have used the ML-POD-ROM \eqref{definition_mixing_length}
\cite[see for example][]{borggaard2008reduced,wang2011two}.
Improvements to the mixing-length model \eqref{definition_mixing_length} in which the EV coefficient is mode
dependent were proposed in \cite{RF94,cazemier1998proper,podvin2009proper}.

\subsubsection{The Smagorinsky POD reduced-order model (S-POD-ROM)}
\label{sss_les_smagorinsky}

A potential improvement over the simplistic mixing-length hypothesis is to replace the constant $\nu_{ML}$ 
in \eqref{ML_POD_ROM_1}-\eqref{ML_POD_ROM_2} (which is computed only once, at the beginning of the 
simulation) with a variable turbulent viscosity (which is recomputed at every time step),
such as that proposed in \cite{Sma63}.
This yields a POD closure model in which the viscosity coefficient 
is increased by 
\begin{eqnarray}
\nu_{S}
:= 2 \, (C_S \, \delta)^2 \, \| \D(\bu_r) \| ,
\label{definition_smagorinsky}
\end{eqnarray}
where $C_S$ is the Smagorinsky constant, $\delta$ is the lengthscale defined in \S\,\ref{ss_pod_lengthscale} 
and $\| \D(\bu_r) \|$ is the Frobenius norm of the deformation tensor $\D(\bu_r)$.
Using the EV ansatz in \eqref{definition_smagorinsky}, the {\em Smagorinsky POD reduced-order model (S-POD-ROM)} 
has the form \eqref{eq:a_closure}, where
\begin{eqnarray}
\widetilde{b}_k({\bf a}) 
&=& -2 \, (C_S \, \delta)^2 \, \left(\nabla {\boldsymbol\phi}_k, \| \md({\bf u}_r) \| \frac{\nabla {\bf U} 
+ \nabla {\bf U}^{T}}{2} \right), \label{S_POD_ROM_1} \\
\widetilde{A}_{km}({\bf a}) 
&=& -2 \, (C_S \, \delta)^2 \, \left(\nabla {\boldsymbol\phi}_k, \| \md({\bf u}_r) \| \frac{\nabla {\boldsymbol\phi}_m 
+ \nabla {{\boldsymbol\phi}_m}^{T}}{2} \right). \label{S_POD_ROM_2}
\end{eqnarray}
The S-POD-ROM \eqref{S_POD_ROM_1}-\eqref{S_POD_ROM_2} was proposed in \cite{borggaard2008reduced} 
\cite[see also][]{noack2002low} and was used in the reduced-order modeling of 
structurally dominated 3D turbulent flows in \cite{wang2011two,ullmann2010pod}.
Its advantage over the ML-POD-ROM \eqref{ML_POD_ROM_1}-\eqref{ML_POD_ROM_2} is obvious: 
the latter utilizes a constant EV coefficient at every time step, whereas the former recomputes the EV coefficient
(which depends on $\| \D(\bu_r) \|$) at every time step.
To address the significant computational burden posed by the recalculation of the Smagorinsky EV coefficient
at every time step, a novel two-level discretization algorithm proposed in \cite{wang2011two} is employed in 
\S\,\ref{s_numerical_results}.

\subsubsection{The Variational Multiscale POD reduced-order model (VMS-POD-ROM)}
\label{sss_les_vms}

The VMS method, a state-of-the-art LES closure modeling strategy, was introduced 
in \cite{HMJ00,HMOW01,HOM01}. 
The VMS method is based on the principle of {\em locality} of energy transfer, i.e., it uses the ansatz that 
energy is transfered mainly between the neighboring scales.
In \cite{CSB03}, the transfer of energy 
among POD modes for turbulent flow past a backward-facing step (a non-homogeneous 
separated flow) was investigated numerically.
In their report, it was shown that the Fourier-decomposition based concepts of energy cascade 
and locality of energy transfer are also valid in the POD context \cite[Figures 3 and 4 
in][]{CSB03}.
Thus, VMS closure models represent a natural choice for POD-ROM.

To develop the VMS POD closure model, we start by decomposing the finite set of POD modes $\bX^r$ into 
the direct sum of \textit{large resolved} POD modes $\bX^r_L$ and \textit{small resolved} POD modes $\bX^r_S$:
\begin{eqnarray}
\bX^r &=& \bX^r_L \oplus \bX^r_S, 
\quad \text{where} \label{les_10} \\
\bX^r_L &:=& \text{span}\left\{\bphi_1, \bvarphi_2, \ldots, \bvarphi_{r_L} \right\} 
\quad \text{and} \label{les_11} \\
\bX^r_S &:=& \text{span}\left\{\bphi_{r_L+1}, \bphi_{r_L+2}, \ldots, \bvarphi_{r} \right\} .
\label{les_12}
\end{eqnarray}
Accordingly, we decompose $\bu_r$ into two components: 
$\bu_r^L$ representing the \textit{large resolved} scales, and
$\bu_r^S$ representing the \textit{small resolved} scales:
\begin{eqnarray}
\bu_r = \bu_r^L + \bu_r^S,
\label{les_13}
\end{eqnarray}
where
\begin{eqnarray}
\bu_r^L
&=& {\bf U} 
+ \sum_{j=1}^{r_L} a_j \, \bphi_k \, ,\\
\bu_r^S 
&=& \sum_{j=r_L+1}^{r} a_j \bphi_k .
\label{les_13_b}
\end{eqnarray}
The two components $\bu_r^L$ and $\bu_r^S$ represent the 
projections of $\bu_r$ onto the two spaces $\bX^r_L$ and 
$\bX^r_S$, respectively.
The general POD-ROM framework \eqref{eq:a_closure} can now be separated 
into two equations - one for $\ba^L$ (the vector of POD coefficients of $\bu_r^L$) 
and one for $\ba^S$ (the vector of POD coefficients of $\bu_r^S$).
The \textit{Variational Multiscale POD reduced-order model (VMS-POD-ROM)} 
applies an eddy viscosity term to the small resolved scales only, following the 
principle of locality of energy transfer.
The VMS-POD-ROM reads:
\begin{eqnarray}
  && \left[ \begin{array}{c} \dot{\bf a}^L \\ \dot{\bf a}^S \end{array} \right] =
  \left[ \begin{array}{c}     {\bf b}^L \\     {\bf b}^S \end{array} \right]
  + {\bf A}^r \left[ \begin{array}{c} {\bf a}^L \\ {\bf a}^S \end{array} \right]
  + \left[ \begin{array}{cc}  {\bf A}^L & {\bf 0} \\    
                                         {\bf 0} & {\bf A}^S + \widetilde{\bf A}^S({\bf a}^S) 
                \end{array} \right]
     \left[ \begin{array}{c} {\bf a}^L \\ {\bf a}^S \end{array} \right]
     \nonumber \\[0.2cm]
  && \hspace*{1.25cm} 
  + \left[ \begin{array}{c} {\bf a}^L \\ {\bf a}^S \end{array} \right]^T           
      {\bf B} \, 
     \left[ \begin{array}{c} {\bf a}^L \\ {\bf a}^S \end{array} \right]    
     \label{VMS_POD_ROM_1}
\end{eqnarray}

The finite dimensional system \eqref{VMS_POD_ROM_1} 
can be written componentwise as follows:
\begin{eqnarray}
\dot{a}_k^L(t) 
&=&  b_k^L
+ \sum_{m=1}^{r} A_{km}^r a_m(t) 
+ \sum_{j=1}^{r_L} A_{kj}^L a_j(t)
+ \sum_{m=1}^r \sum_{n=1}^r B_{kmn}a_n(t)a_m(t) ,
\label{VMS_POD_ROM_3} \\
&& \hspace*{7.5cm}
\forall \, k = 1, \ldots, r_L , 
\nonumber \\[0.1cm]
\dot{a}_k^S(t) 
&=&  b_k^S
+ \sum_{m=1}^{r} A_{km}^r a_m(t) 
+ \sum_{j=r_L+1}^{r}  \left( A_{kj}^S + \widetilde{A}_{kj}^S \right) a_j(t)
\label{VMS_POD_ROM_4} \\
&& \hspace*{0.45cm} + \sum_{m=1}^r \sum_{n=1}^r B_{kmn}a_n(t)a_m(t) 
\hspace*{3.0cm}
\forall \, k = r_L + 1, \ldots, r,
\nonumber
\end{eqnarray}
where
\begin{eqnarray}
b_k^L 
&=& -\left( \boldsymbol\phi_k, {\bf U} \cdot \nabla {\bf U} \right) 
- \frac{2}{\mbox{Re}} \left( \nabla \boldsymbol\phi_k, \frac{\nabla {\bf U} +\nabla {\bf U}^{T}}{2} \right), 
\label{VMS_POD_ROM_3}\\
A_{km}^{r} 
&=& -(\boldsymbol\phi_k, {\bf U} \cdot \nabla \boldsymbol\phi_m) 
- (\boldsymbol\phi_k, \boldsymbol\phi_m \cdot \nabla {{\bf U}}), 
\label{VMS_POD_ROM_5} \\[0.2cm]
A_{kj}^{L} 
&=& - \frac{2}{\mbox{Re}} \left( \nabla \boldsymbol\phi_k, 
\frac{\nabla \boldsymbol\phi_j +\nabla \boldsymbol\phi_j^{T}}{2} \right), 
\label{VMS_POD_ROM_6}\\[0.1cm]
B_{kmn}
&=& -(\boldsymbol\phi_k, \boldsymbol\phi_m \cdot \nabla \boldsymbol\phi_n) , 
\label{VMS_POD_ROM_6b} \\[0.3cm]
b_k^S 
&=& -\left( \boldsymbol\phi_k, {\bf U} \cdot \nabla {\bf U} \right), 
\label{VMS_POD_ROM_7}\\[0.2cm]
A_{kj}^{S} 
&=& - \frac{2}{\mbox{Re}} \left( \nabla \boldsymbol\phi_k, 
\frac{\nabla \boldsymbol\phi_j +\nabla \boldsymbol\phi_j^{T}}{2} \right), 
\label{VMS_POD_ROM_8}\\
\widetilde{A}_{kj}^{S}({\bf a}) 
&=& -2 \, (C_S \, \delta)^2 \, \left(\nabla {\boldsymbol\phi}_k, \| \md({\bf u}_r^S + {\bf U}) \| 
\frac{\nabla {\boldsymbol\phi}_j + \nabla {{\boldsymbol\phi}_j}^{T}}{2} \right). 
\label{VMS_POD_ROM_9} 
\end{eqnarray}
We emphasize that the system of equations \eqref{VMS_POD_ROM_1}
is {\em coupled} through two terms:
(i) ${\bf a}^T {\bf B} {\bf a}$, which represents the nonlinearity $(\bu^{r} \cdot \nabla) \bu^{r}$; and
(ii) ${\bf A}^{r} {\bf a}$, which represents the term $(\bu^{r} \cdot \nabla) \bu^{r}$ linearized around
the centering trajectory ${\bf U}$. 
The difference between the VMS-POD-ROM \eqref{VMS_POD_ROM_1}-\eqref{VMS_POD_ROM_9} 
and the S-POD-ROM \eqref{S_POD_ROM_1}-\eqref{S_POD_ROM_2} is that the former acts only on 
the small resolved scales (since the Smagorinsky EV term 
$(C_S \, \delta)^2 \, \| \md({\bf u}_r^S + {\bf U}) \|$ is included only in 
the equation corresponding to ${\bf a}^S$), whereas the latter acts on all (both large and small)
resolved scales.

The VMS-POD-ROM \eqref{VMS_POD_ROM_1}-\eqref{VMS_POD_ROM_9} was announced in
\cite{borggaard2008reduced}.
This study, however, represents its first careful derivation and through investigation in the 
numerical simulation of a 3D turbulent flow.
We note that a fundamentally different VMS LES closure model that utilizes the NSE residual 
was proposed in \cite{bazilevs2007variational}; this model was used in a POD setting in 
\cite{bergmann2009enablers}.
Yet another VMS-POD-ROM, inspired from the numerical stabilization methods developed in
\cite{layton2002connection,guermond1999stabilization,john2005finite,john2010variational}, was
proposed, analyzed and tested in \cite{iliescu2010variational}.
We emphasize that the VMS-POD-ROM \eqref{VMS_POD_ROM_1}-\eqref{VMS_POD_ROM_9} is different from both the model used in \cite{bergmann2009enablers}
and that used in \cite{iliescu2010variational}.

\subsubsection{Dynamic Subgrid-Scale POD reduced-order model (DS-POD-ROM)}
\label{sss_dynamic}

For all three POD-ROM closure models defined up to this point 
(i.e., ML-POD-ROM \eqref{ML_POD_ROM_1}-\eqref{ML_POD_ROM_2},
S-POD-ROM \eqref{S_POD_ROM_1}-\eqref{S_POD_ROM_2}, 
and VMS-POD-ROM \eqref{VMS_POD_ROM_1}-\eqref{VMS_POD_ROM_9}),
the definition has been entirely phenomenological.
Indeed, arguing that the role of the discarded POD modes is to extract energy from the system, we used
an EV ansatz to derive closure models of increasing complexity and physical accuracy.
The {\em dynamic subgrid-scale (DS)} POD-ROM closure model is also of EV type.
Its derivation, however, needs a precise definition of the filtering operation.
The DS closure model has its origins in LES, where it is considered state-of-the-art \cite[see for example][]{Sag06}.
In LES, the filtering operation is effected by convolving the flow variables with a rapidly decaying spatial 
filter.
In POD, the filtering operation is effected by using the POD Galerkin projection described in \S 
\ref{s_pod_filtering} (see \eqref{galerkin_projection}).
To derive the precise POD filtered equations, we start with the NSE \eqref{NSE} in which the velocity
$\bu$ is replaced by its POD approximation 
${\bf u}({\bf x},t) \approx {\bf u}_r({\bf x},t) \equiv {\bf U}({\bf x}) + \sum_{j=1}^r a_j(t) \boldsymbol\phi_j({\bf x})$ 
in \eqref{eq:yr}, and obtain 
\begin{eqnarray}
\frac{\partial \bu_r}{\partial t}
- Re^{-1} \Delta \bu_r
+ ( \bu_r \cdot \nabla) \, \bu_r
+ \nabla p
= 0 .
\label{sss_dynamic_1}
\end{eqnarray}
Using the fact that $\nabla \cdot \bu_r = 0$ in \eqref{sss_dynamic_1}, we get 
$( \bu_r \cdot \nabla) \, \bu_r = \nabla \cdot (\bu_r \, \bu_r)$.
Thus, \eqref{sss_dynamic_1} can be rewritten as
\begin{eqnarray}
\frac{\partial \bu_r}{\partial t}
- Re^{-1} \Delta \bu_r
+ \nabla \cdot (\bu_r \, \bu_r)
+ \nabla p
= 0 .
\label{sss_dynamic_2}
\end{eqnarray}
Applying the POD filtering operation \eqref{galerkin_projection} to \eqref{sss_dynamic_2}, using the fact 
that the POD Galerkin projection is a linear operator, and {\em assuming} that differentiation and POD 
filtering commute, we obtain
\begin{eqnarray}
\frac{\partial \obu_r}{\partial t}
- Re^{-1} \Delta \obu_r
+ \nabla \cdot ({\overline{\bu_r \, \bu_r}})
+ \nabla \op
= 0 .
\label{sss_dynamic_3}
\end{eqnarray}
We note that, if filtering and differentiation do not commute, one has to estimate the commutation error \cite[see for example][]{VG2004,VLM98,BIL05}.
We also note that, since the POD filtering operation is the Galerkin projection \eqref{galerkin_projection},  
$\obu_r = \bu_r$.
For consistency with the nonlinear term notation, we still use the $\obu_r$ notation in what follows.

The POD filtered equation \eqref{sss_dynamic_3} can be rewritten as
\begin{eqnarray}
\frac{\partial \obu_r}{\partial t}
- Re^{-1} \Delta \obu_r
+ \nabla \cdot (\obu_r \, \obu_r)
+ \nabla \cdot (\btau_r)
+ \nabla \op
= 0 ,
\label{sss_dynamic_4}
\end{eqnarray}
where
\begin{eqnarray}
\btau_r
= {\overline{\bu_r \, \bu_r}} - \obu_r \, \obu_r
\label{sss_dynamic_5}
\end{eqnarray}
is the POD subfilter-scale stress tensor.
Thus, the POD-G-ROM \eqref{pod_g} amounts to setting $\btau_r = 0$.
For turbulent flows, as we have already mentioned, this approximation is flawed.
Thus, one needs to address the POD closure problem, i.e., to model the POD sufilter-scale stress tensor
$\btau_r$ in terms of POD filtered velocity $\obu_r$.
We note that the POD closure problem is exactly the LES closure problem, in which the spatial filtering is 
replaced by POD Galerkin projection.
For all three POD-ROM closure models defined so far in this section 
(i.e., ML-POD-ROM \eqref{ML_POD_ROM_1}-\eqref{ML_POD_ROM_2},
S-POD-ROM \eqref{S_POD_ROM_1}-\eqref{S_POD_ROM_2}, 
and VMS-POD-ROM \eqref{VMS_POD_ROM_1}-\eqref{VMS_POD_ROM_9}),
the closure
problem has been addressed by assuming an EV ansatz for $\btau_r$.
The DS-POD-ROM employs an EV ansatz as well; specifically, the Smagorinsky model is utilized:
\begin{eqnarray}
\btau_r
:= (C_S \, \delta)^2 \, \| \D(\bu_r) \| ,
\label{sss_dynamic_6}
\end{eqnarray}
in which $C_S$ is not a constant (as in the Smagorinsky model), but a function of space and time, i.e., 
$C_S = C_S({\bf x}, t)$.
To compute $C_S({\bf x}, t)$, we follow the LES derivation in \cite{Sag06} and replace the LES spatial
filtering with the POD Galerkin projection.
Since there are two spatial filters in the LES derivation of the DS model, we define a second POD Galerkin
projection (in addition to that defined in \eqref{galerkin_projection}):
For all ${\bf u} \in \bX$, the second (test) Galerkin projection $\tbu \in \bX^R$ (where $R < r$) is the 
solution of the following equation:
\begin{eqnarray}
(\bu - \tbu , \bphi) = 0
\qquad \forall \, \bphi \in \bX^R.
\label{galerkin_projection_2}
\end{eqnarray}
Applying the second POD filtering operation \eqref{galerkin_projection_2} to \eqref{sss_dynamic_3}, we obtain:
\begin{eqnarray}
\frac{\partial \tobu_r}{\partial t}
- Re^{-1} \Delta \tobu_r
+ \nabla \cdot (\tobu_r \, \tobu_r)
+ \nabla \cdot (\bT_r)
+ \nabla {\widetilde \op}
= 0 ,
\label{sss_dynamic_7}
\end{eqnarray}
where
\begin{eqnarray}
\bT_r
= {\widetilde {\overline{\bu_r \, \bu_r}}} - \tobu_r \, \tobu_r
\label{sss_dynamic_8}
\end{eqnarray}
is the second POD sufilter-scale stress tensor.
We note that the following identity (called the ``Germano identity" in LES) holds:
\begin{eqnarray}
\bT_r
= {\widetilde {\overline{\bu_r \, \bu_r}}} - \tobu_r \, \tobu_r
= \left( {\widetilde{\obu_r \, \obu_r} } - \tobu_r \, \tobu_r \right)
+ \left( {\widetilde {\overline{\bu_r \, \bu_r}}} - {\widetilde{\obu_r \, \obu_r} } \right) 
= \bL_r  + \widetilde{\btau_r} ,
\label{sss_dynamic_9}
\end{eqnarray}
where $\bL_r = \widetilde{\obu_r \, \obu_r}  - \tobu_r \, \tobu_r$ and 
$\widetilde{\btau_r} = {\widetilde {\overline{\bu_r \, \bu_r}}} - {\widetilde{\obu_r \, \obu_r} }$.
We {\em assume the same EV ansatz} for the two POD subfilter-scale stress tensors, 
$\btau_r$ and $\bT_r$:
\begin{eqnarray}
\bT_r
&\approx& - 2 \, (C_S \, \widetilde{\delta})^2 \, \| \D(\tobu_r) \| \,  \D(\tobu_r) \label{sss_dynamic_10} \\
\btau_r
&\approx& - 2 \, (C_S \, \delta)^2 \, \| \D(\obu_r) \| \,  \D(\obu_r) ,\label{sss_dynamic_11}
\end{eqnarray}
where $\widetilde{\delta}$ is the filter radius used in the second POD filtering operation 
\eqref{galerkin_projection_2}.
{\em Assuming} that $C_S$ remains constant under the second POD filtering 
\eqref{galerkin_projection_2}, we obtain:
\begin{eqnarray}
\widetilde{\btau_r}
\approx \widetilde{- 2 \, (C_S \, \delta)^2 \, \| \D(\obu_r) \| \,  \D(\obu_r)} 
\approx - 2 \, (C_S \, \delta)^2 \, \widetilde{\| \D(\obu_r) \| \,  \D(\obu_r)} .
\label{sss_dynamic_12}
\end{eqnarray}
Plugging \eqref{sss_dynamic_10} and \eqref{sss_dynamic_12} into \eqref{sss_dynamic_9} 
we obtain:
\begin{eqnarray}
- 2 \, (C_S \, \widetilde{\delta})^2 \, \| \D(\tobu_r) \| \,  \D(\tobu_r)
= \left( \widetilde{\obu_r \, \obu_r}  - \tobu_r \, \tobu_r \right)
- 2 \, (C_S \, \delta)^2 \, \widetilde{\| \D(\obu_r) \| \,  \D(\obu_r)} .
\label{sss_dynamic_13}
\end{eqnarray}
We note that $C_S$ is the only unknown in \eqref{sss_dynamic_13}, all the other terms 
being computable quantities.
Since all the terms in \eqref{sss_dynamic_13} are tensors, the unknown $C_S$ cannot
satisfy all nine equations.
Thus, the following least squares approach is considered instead:
\begin{eqnarray}
&\stackrel{\text{\small min}}{C_S^2}&
\left[ 
\left( \widetilde{\obu_r \, \obu_r}  - \tobu_r \, \tobu_r \right)
- 2 \, (C_S \, \delta)^2 \, \widetilde{\| \D(\obu_r) \| \,  \D(\obu_r)} 
+ 2 \, (C_S \, \widetilde{\delta})^2 \, \| \D(\tobu_r) \| \,  \D(\tobu_r)                  
\right]
: \nonumber \\
&&\left[
\left( \widetilde{\obu_r \, \obu_r}  - \tobu_r \, \tobu_r \right)
- 2 \, (C_S \, \delta)^2 \, \widetilde{\| \D(\obu_r) \| \,  \D(\obu_r)} 
+ 2 \, (C_S \, \widetilde{\delta})^2 \, \| \D(\tobu_r) \| \,  \D(\tobu_r)                  
\right] .
\label{sss_dynamic_14}
\end{eqnarray}
The solution $C_S(\bx, t)$ of \eqref{sss_dynamic_14} is:
\begin{eqnarray}
&& \hspace*{-0.6cm} C_S^2(\bx, t) = \label{sss_dynamic_15} \\
&& \hspace*{-0.6cm} \frac{\left[ {\widetilde{\obu_r \, \obu_r} } - \tobu_r \, \tobu_r \right] 
\colon \left[ 2 \, \delta^2 \, \widetilde{\| \D(\obu_r) \| \,  \D(\obu_r)} - 2 \, \widetilde{\delta}^2 \, \| \D(\tobu_r) \| \,  \D(\tobu_r) \right] }
{\left[ 2 \, \delta^2 \, \widetilde{\| \D(\obu_r) \| \,  \D(\obu_r)} - 2 \, \widetilde{\delta}^2 \, \| \D(\tobu_r) \| \,  \D(\tobu_r) \right]
\colon \left[ 2 \, \delta^2 \, \widetilde{\| \D(\obu_r) \| \,  \D(\obu_r)} - 2 \, \widetilde{\delta}^2 \, \| \D(\tobu_r) \| \,  \D(\tobu_r) \right] } . \nonumber 
\end{eqnarray}
Since the stress tensors can be computed directly from the resolved field, \eqref{sss_dynamic_15} 
yields a time- and space-dependent formula for $C_S(\bx,t)$.

Thus, the DS-POD-ROM increases the viscosity 
coefficient by 
\begin{eqnarray}
\nu_{DS}
:= 2 \, \bigl(C_S(\bx, t) \, \delta \bigr)^2 \, \| \D(\bu_r) \| ,
\label{definition_dynamic_sgs}
\end{eqnarray}
where $C_S(\bx, t)$ is the coefficient in \eqref{sss_dynamic_15}, $\delta$ is the lengthscale defined 
in \S\,\ref{ss_pod_lengthscale} and $\| \D(\bu_r) \|$ the Frobenius norm of the deformation tensor 
$\D(\bu_r)$.
Thus, the {\em Dynamic Subgrid-Scale POD reduced-order model (DS-POD-ROM)} has the form \eqref{eq:a_closure}, where
\begin{eqnarray}
\widetilde{b}_k({\bf a}) 
&=& -2 \, \delta^2 \, \left(\nabla {\boldsymbol\phi}_k, C_S^2(\bx, t) \,  \, \| \md({\bf u}_r) \| \frac{\nabla {\bf U} +\nabla {\bf U}^{T}}{2} \right), \label{DS_POD_ROM_1} \\
\widetilde{A}_{km}({\bf a}) 
&=& -2 \, \delta^2 \, \left(\nabla {\boldsymbol\phi}_k, C_S^2(\bx, t) \, \| \md({\bf u}_r) \| \frac{\nabla {\boldsymbol\phi}_m + \nabla {{\boldsymbol\phi}_m}^{T}}{2} \right). \label{DS_POD_ROM_2}
\end{eqnarray}

Note that $\nu_{DS}$ defined in~\eqref{definition_dynamic_sgs} can take negative values.
This can be interpreted as \textit{backscatter}, the inverse transfer of energy from high index 
POD modes to low index ones.
The notion of backscatter, well-established in LES \cite[see for example][]{Sag06}, was also found in a 
POD setting in the numerical investigation in \cite{CSB03}.

\section{Numerical tests}
\label{s_numerical_results}

In this section, we use a structurally dominated 3D turbulent flow problem to test the four POD-ROMs 
described in \S\,\ref{s_closure_models}:
(i) the ML-POD-ROM \eqref{ML_POD_ROM_1}-\eqref{ML_POD_ROM_2}; 
(ii) the S-POD-ROM \eqref{S_POD_ROM_1}-\eqref{S_POD_ROM_2};
(iii) the new VMS-POD-ROM \eqref{VMS_POD_ROM_1}-\eqref{VMS_POD_ROM_9}; and
(iv) the new DS-POD-ROM \eqref{DS_POD_ROM_1}-\eqref{DS_POD_ROM_2}.
We also include results for the POD-G-ROM \eqref{pod_g} (i.e., a POD-ROM without any
closure model).
A successful POD closure model should at least perform better than the POD-G-ROM \eqref{pod_g}.
Finally, a DNS projection of the evolution of the POD modes served as benchmark for our numerical 
simulations:
The closeness to the DNS data was used as a criterion for the success of the POD closure model.
The qualitative behavior of all POD-ROMs is judged according to the following two criteria: 
(i) the kinetic energy spectrum,  which represents the temporal average behavior of the POD-ROMs; and
(ii) the time evolution of the POD coefficients, which measures the instantaneous behavior of the POD-ROMs. 
In \S\,\ref{ss_numerical_methods}, details of the numerical methods and parameter choices are given. 
In \S\,\ref{ss_numerical_results}, numerical results are presented and discussed. 

\subsection{Numerical Methods and Parameter Choices}
\label{ss_numerical_methods}
We investigate all four POD-ROMs in the numerical simulation of 3D flow past 
a circular cylinder at $\mbox{Re} = 1,000$. 
The wake of the flow is fully turbulent. 
A parallel fluid flow solver is employed on a $144\times 192\times 16$ finite volume mesh 
on the time interval $[0,300]$ to generate the DNS data. 
For details on numerical discretization, the reader is referred to Appendix A in \cite{wang2011two}. 

Collecting $1,000$ snapshots of the velocity field ($u_1,u_2,u_3$) 
over the time interval $[0,75]$
and applying the method of snapshots developed in \cite{Sir87abc}, we obtain the POD basis 
$\{ {\boldsymbol \phi}_1, \cdots, {\boldsymbol \phi}_N \}$. 
These POD modes are then interpolated onto a structured quadratic finite element mesh 
with nodes coinciding with the nodes used in the original DNS finite volume discretization. 
The first $r=6$ POD modes capture 84\% the system's kinetic energy. 
These modes are used in all POD-ROMs that we investigate next.
For all the POD-ROMs, the time discretization was effected by using the explicit Euler method 
with $\Delta t= 7.5\times10^{-4}$. 

It is important to note that the quadratic nonlinearity in the NSE (\ref{NSE}) allows for easy 
precomputation of the vector $\bf b$, the matrix $\bf A$ and the tensor $\bf B$ in the POD-G-ROM \eqref{eq:a}. 
For the general nonlinear EV POD closure model \eqref{eq:a_closure}, however, the vector 
$\widetilde{\bf b}({\bf a})$ and the matrix $\widetilde{\bf A}({\bf a})$ that correspond to the 
additional closure terms have to be recomputed (reassembled) at each time step. 
Since the POD basis functions are global, although only a few are used in POD-ROMs ($r \ll N$), 
reassembling $\widetilde{\bf b}({\bf a})$ and $\widetilde{\bf A}({\bf a})$ at each time step would 
dramatically increase the CPU time of the corresponding POD-ROM. 
Thus, the major advantage of POD-ROMs (the dramatic decrease of computational time), would
be completely lost.

To ensure a high computational efficiency of the POD-ROMs, we utilize two approaches:
(1) Instead of updating the closure terms in the POD-ROMs every time step, we recompute 
them every $1.5$ time units (i.e., every $20,000$ time steps). 
The previous numerical investigations in \cite{wang2011two} showed that this approach does not
compromise the numerical accuracy of the S-POD-ROM \eqref{S_POD_ROM_1}-\eqref{S_POD_ROM_2}.
(2) We employ the two-level algorithm introduced in \cite{wang2011two} to discretize the 
nonlinear closure models. 
Before briefly describing the two-level algorithm, we emphasize that, in order to maintain a
fair numerical comparison of the four POD-ROMs, we used both algorithmic choices
listed above in {\em all} four POD-ROMs.
Therefore, the success or failure of the POD-ROM can solely be attributed to the closure
term, which is the only distinguishing feature among all POD-ROMs, and not to the specific
algorithmic choices, which are the same for all POD-ROMs.

The two-level algorithm used in all four POD-ROMs is summarized below.
\begin{center}\hspace*{-3.0cm}
\framebox[4.2in][t]{
\begin{minipage}[c]{4.0in}
\  \\[-0.25in]
\begin{eqnarray}
&& \ell=0; {\tt compute } \ {\bf b, A, B} \ \text{\tt on the \underline{fine} mesh }; \nonumber \\
&& \text{\tt for } \ell=0 \text{ \tt  to } M-1 \nonumber \\
&& \quad {\tt compute }   \ \widetilde{\bf b}({\bf a}^{\ell}), \, \widetilde{\bf A}({\bf a}^{\ell}) \ 
\text{\tt on the \underline{coarse} mesh }\ \  \label{h2l_algorithm} \\
&& \quad {\bf a}^{\ell+1} := \widetilde{F}({\bf a}^{\ell}); \nonumber \\
&& \text{\tt endfor} \nonumber 
\end{eqnarray}
\ \\[-0.35in]
\end{minipage}\hspace*{0.4in}
\begin{minipage}[c]{2.0in}
{\tt two-level} \\
{\tt algorithm}
\end{minipage}
}
\end{center}
In \eqref{h2l_algorithm}, $M$ represents the total number of time steps. 
The idea in the two-level algorithm is straightforward:
Instead of computing the closure terms $\widetilde{\bf b}({\bf a}^{\ell}), \, \widetilde{\bf A}({\bf a}^{\ell})$ 
directly on the fine mesh (as done in the standard one-level algorithm), the two-level algorithm 
discretizes them on a {\em coarser mesh}. 
Thus, the two-level algorithm is much more efficient than the standard one-level algorithm.
Indeed, in \cite{wang2011two} it was shown that the two-level algorithm \eqref{h2l_algorithm} 
achieves the same level of accuracy as the one-level algorithm while decreasing the computational 
cost by an order of magnitude. 
In all four POD-ROMs, we apply the two-level algorithm with a coarsening factor $R_c=4$ in both radial 
and azimuthal directions. 
Thus, the vectors and matrices related to the nonlinear closure terms are computed on a coarse 
finite element mesh with $37\times 49\times 17$ grid points.

%

In \S\,\ref{ss_pod_lengthscale}, we proposed two definitions for the POD lengthscale $\delta$.
Since in the finite element discretization that we employ, definition \eqref{delta_HLB96} is harder to 
implement than \eqref{delta_AHLS88}, we use the latter. 
Thus, using definition \eqref{delta_AHLS88} with $r=6$, we obtain $\delta=0.1179$, which is the 
POD lengthscale that we will use in all four POD-ROMs.
For the DS-POD-ROM \eqref{DS_POD_ROM_1}-\eqref{DS_POD_ROM_2}, we need to define 
the second (test) Galerkin projection \eqref{galerkin_projection_2} and the corresponding filter
radius $\widetilde{\delta}$.
Choosing $R=1$ in \eqref{galerkin_projection_2} and using \eqref{delta_AHLS88}, we obtain
$\widetilde{\delta}=0.1769$. 

The constants in EV LES models are determined in a straightforward fashion, utilizing scaling 
laws satisfied by general 3D turbulent flows \cite[see for example][]{Sag06}.
Although the energy cascade concept in a POD context was verified numerically in \cite{CSB03},
there are no general scaling laws available in this setting.
Thus, to our knowledge, the correct values for the EV constants $\alpha$ in 
the ML-POD-ROM \eqref{ML_POD_ROM_1}-\eqref{ML_POD_ROM_2} 
and $C_S$ in 
the S-POD-ROM \eqref{S_POD_ROM_1}-\eqref{S_POD_ROM_2} and 
the new VMS-POD-ROM \eqref{VMS_POD_ROM_1}-\eqref{VMS_POD_ROM_9}
are still not known.
To determine these EV constants, we run the corresponding POD-ROM on the short time interval
$[0,15]$ with several different values for the EV constants and choose the value that yields the results
that are closest to the DNS results.
This approach yields the following values for the EV constants: 
$\alpha= 3 \times 10^{-3}$ for the ML-POD-ROM, 
$C_S=0.1426$ for the S-POD-ROM, and
$C_S=0.1897$ for the VMS-POD-ROM.
We emphasize that these EV constant values are optimal only on the short time interval tested, 
and they might actually be non-optimal on the entire time interval $[0,300]$ on which the POD-ROMs 
are tested.
Thus, this heuristic procedure ensures some fairness in the numerical comparison of the four POD-ROMs.

In the VMS-POD-ROM, only the first POD basis is considered as the large resolved POD mode, 
that is, $r_L=1$ in \eqref{les_11}.  
In the DS-POD-ROM, since $\nu_{DS}$ can be negative,  we use a standard ``clipping" procedure
to ensure the numerical stability of the discretization \cite[see for example][]{Sag06}.
Specifically, we let $C_S({\bf x},t) = \max\{C_S({\bf x},t), -0.2\}$.  
The value $-0.2$ is 
determined as follows:
We first run the DS-POD-ROM without ``clipping" for the time interval $[0,15]$ and 
record $C_{S, ave}^{-}$, the average negative value of $C_S({\bf x}, t)$.
We then run on the entire time interval $[0,300]$ the DS-POD-ROM with a ``clipping" 
value $C_{S, ave}^{-} / 2 = -0.2$.
We note that there are alternative procedures to deal with the same issue in LES, such as VDSMwc \cite[]{morinishi2002vector}.  
We utilized the standard ``clipping" procedure described above as a first step in the numerical investigation
of the DS-POD-ROM.

 
 \begin{figure}
\centering
\caption{
(Continued on next page.)
Snapshots of horizontal velocity at $t= 142.5\,s$ for: 
(a) DNS; 
(b) the POD-G-ROM \eqref{pod_g};
(c) the ML-POD-ROM \eqref{ML_POD_ROM_1}-\eqref{ML_POD_ROM_2}; 
(d) the S-POD-ROM \eqref{S_POD_ROM_1}-\eqref{S_POD_ROM_2};
(e) the new VMS-POD-ROM \eqref{VMS_POD_ROM_1}-\eqref{VMS_POD_ROM_9}; and
(f) the new DS-POD-ROM \eqref{DS_POD_ROM_1}-\eqref{DS_POD_ROM_2}.
Five isosurfaces are plotted. 
Note that the POD-G-ROM adds unphysical structures, whereas the ML-POD-ROM eliminates 
some of the DNS structures.
The S-POD-ROM, VMS-POD-ROM, and DS-POD-ROM perform well, capturing a similar amount
of structure as the DNS.
}\label{fig_3d_snapshots} \vspace*{0.1cm}
\begin{minipage}[h]{0.03\linewidth} (a) \end{minipage}
\begin{minipage}[h]{0.8\linewidth} \includegraphics[width=1\textwidth]{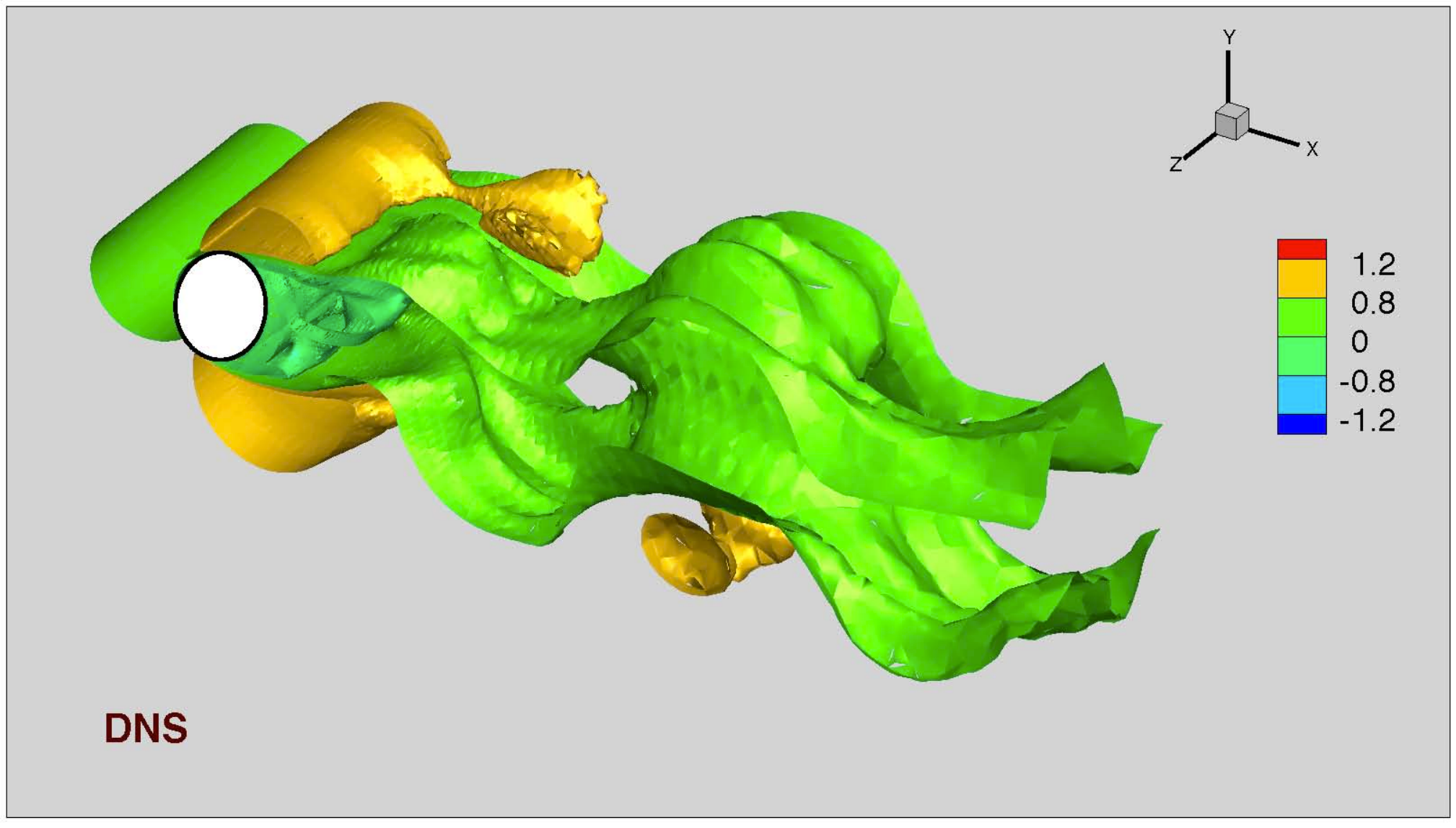}\end{minipage}\\
\begin{minipage}[h]{0.03\linewidth} (b) \end{minipage}
\begin{minipage}[h]{0.8\linewidth} \includegraphics[width=1\textwidth]{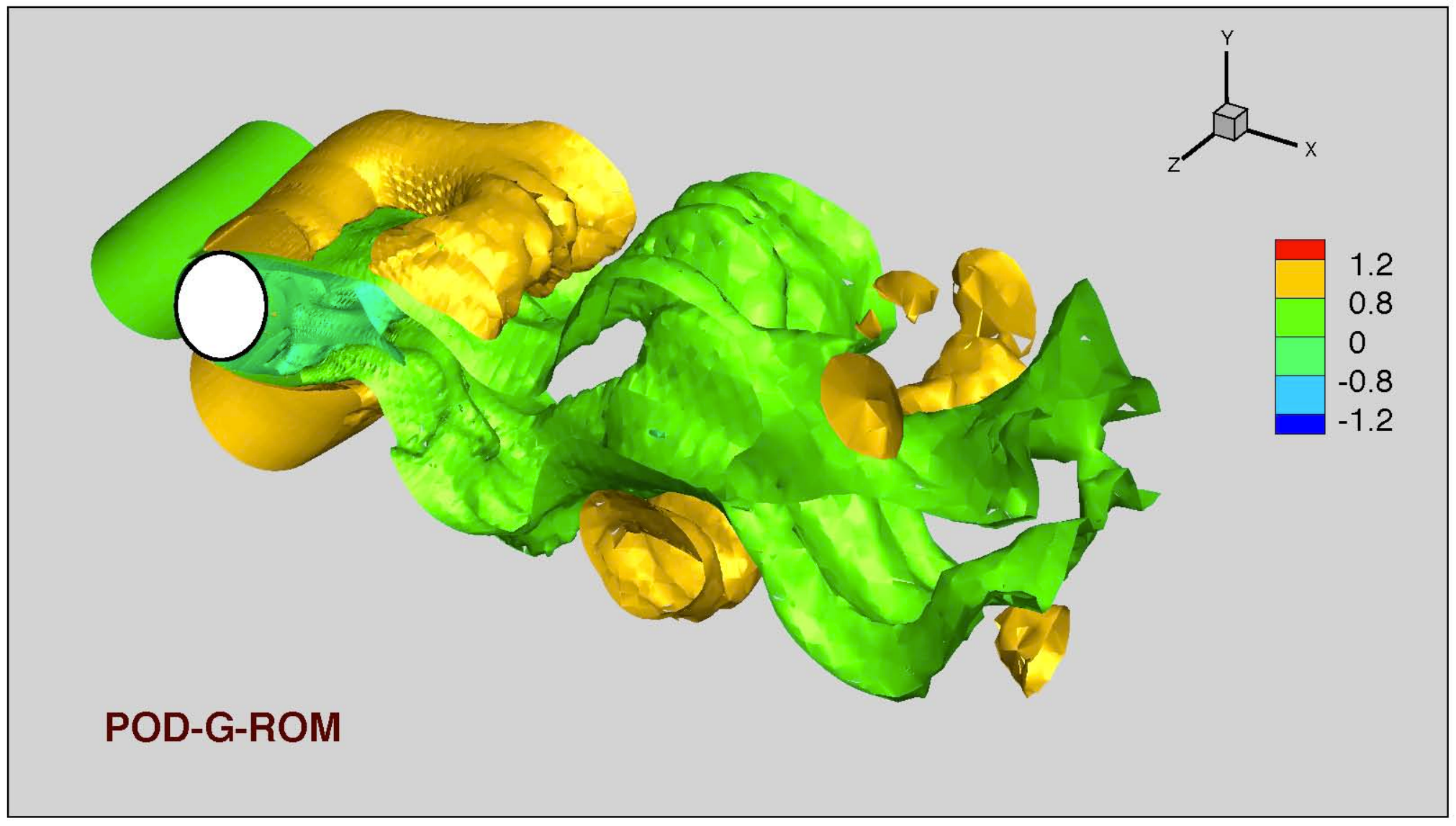}\end{minipage}\\
\begin{minipage}[h]{0.03\linewidth} (c) \end{minipage}
\begin{minipage}[h]{0.8\linewidth} \includegraphics[width=1\textwidth]{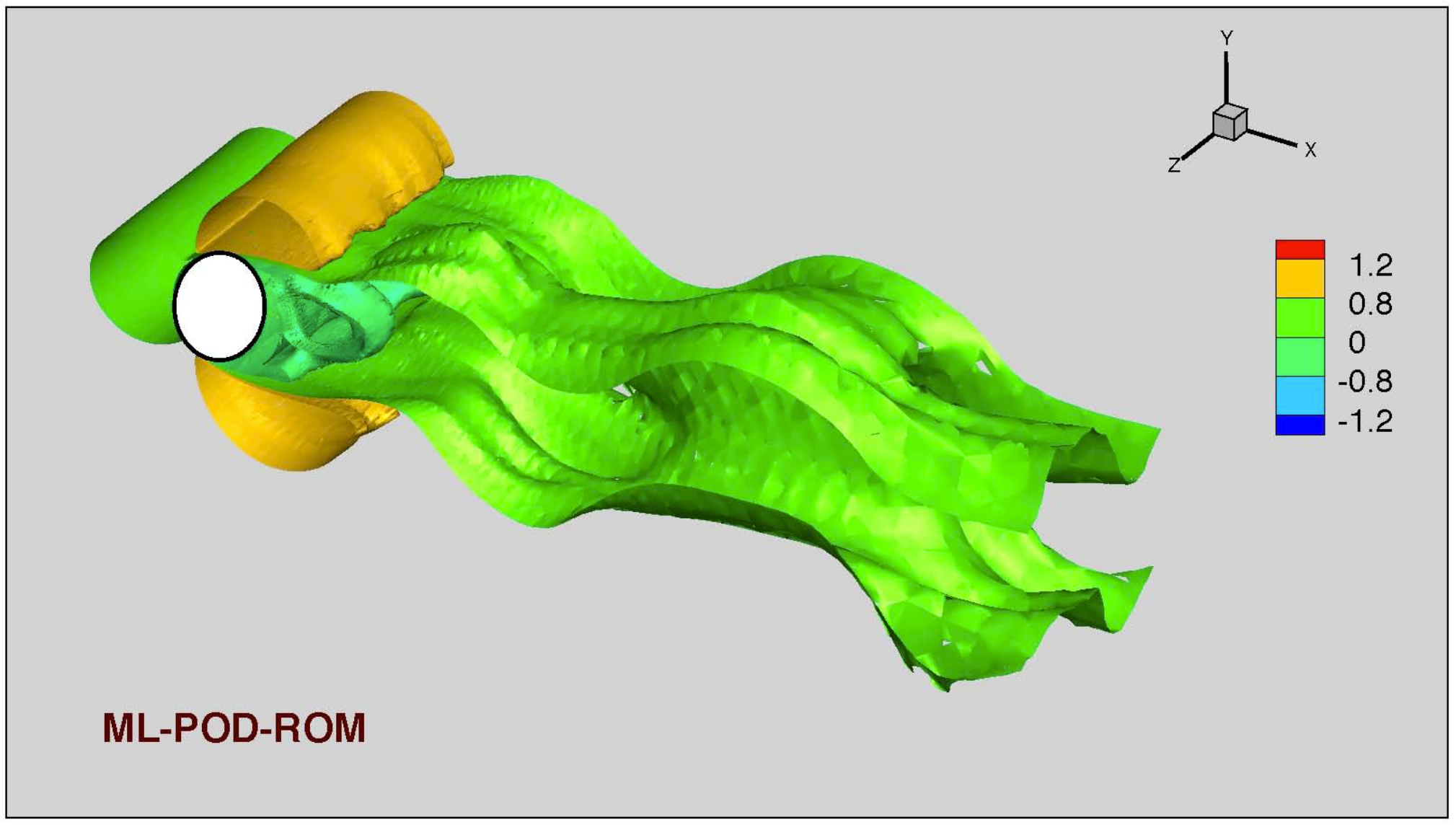}\end{minipage}
\end{figure}

\begin{figure}
\centering
\begin{minipage}[h]{0.03\linewidth} (d) \end{minipage}
\begin{minipage}[h]{0.8\linewidth} \includegraphics[width=1\textwidth]{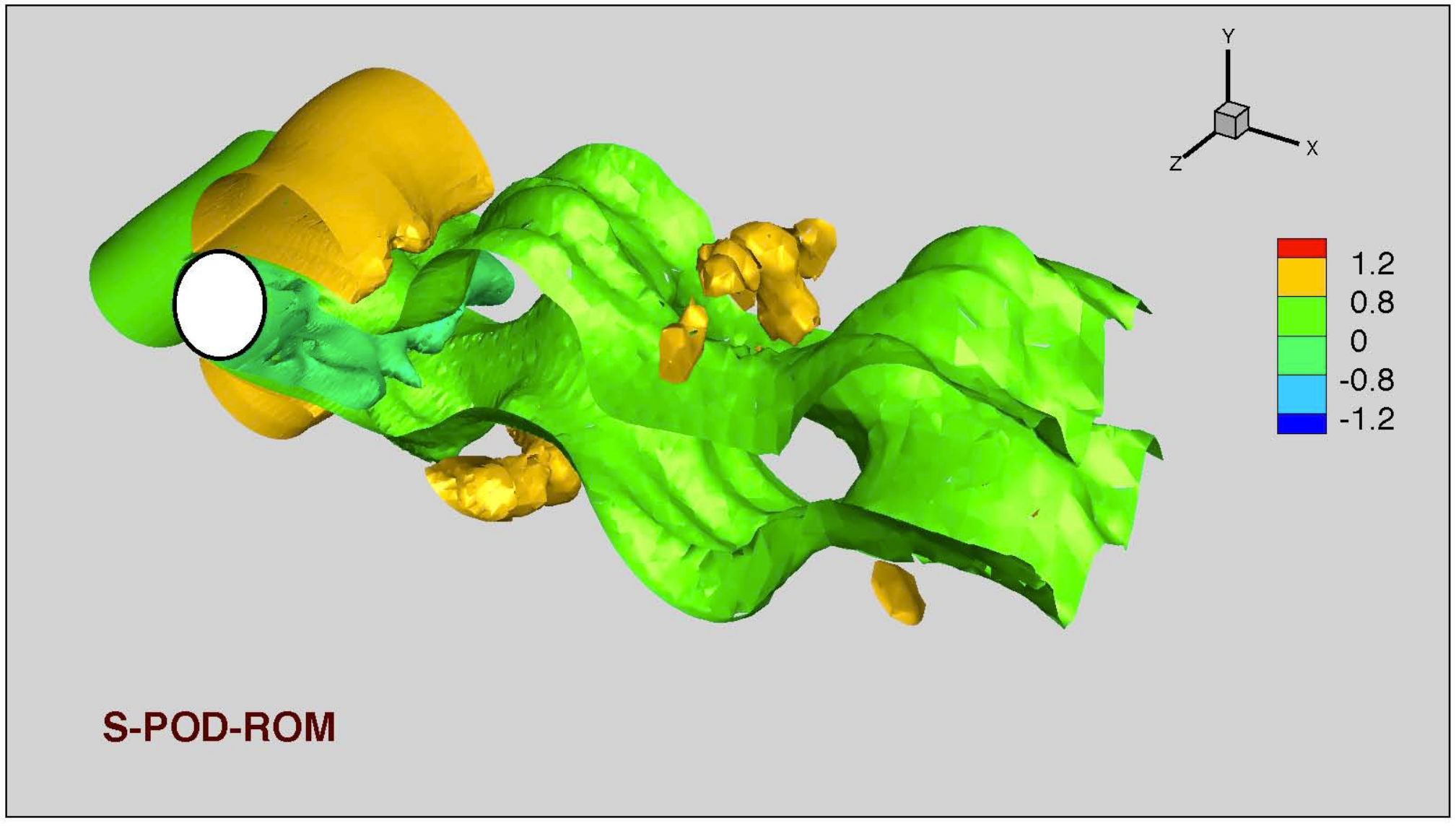}\end{minipage}\\
\begin{minipage}[h]{0.03\linewidth} (e) \end{minipage}
\begin{minipage}[h]{0.8\linewidth} \includegraphics[width=1\textwidth]{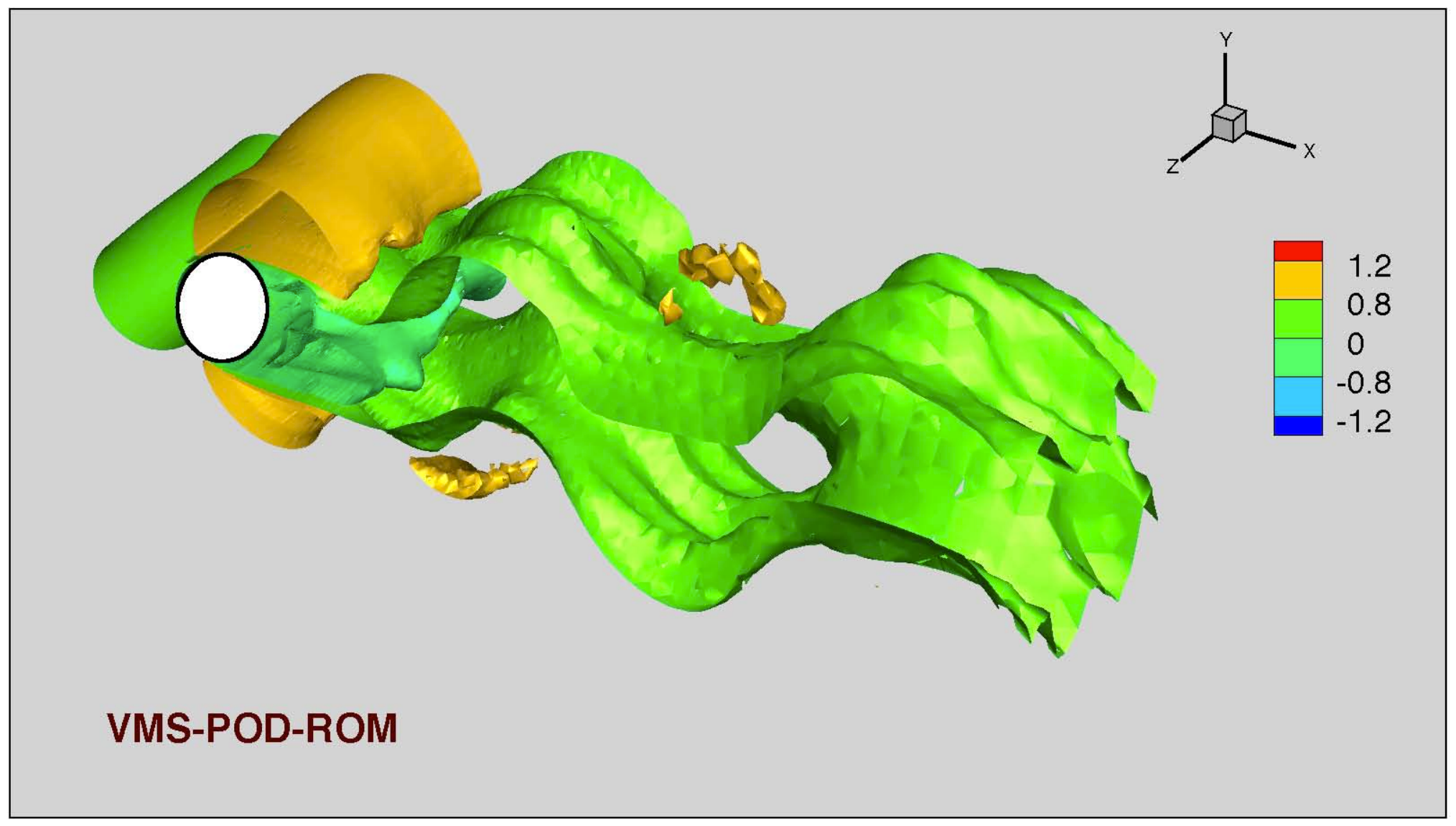}\end{minipage}\\
\begin{minipage}[h]{0.03\linewidth} (f) \end{minipage}
\begin{minipage}[h]{0.8\linewidth} \includegraphics[width=1\textwidth]{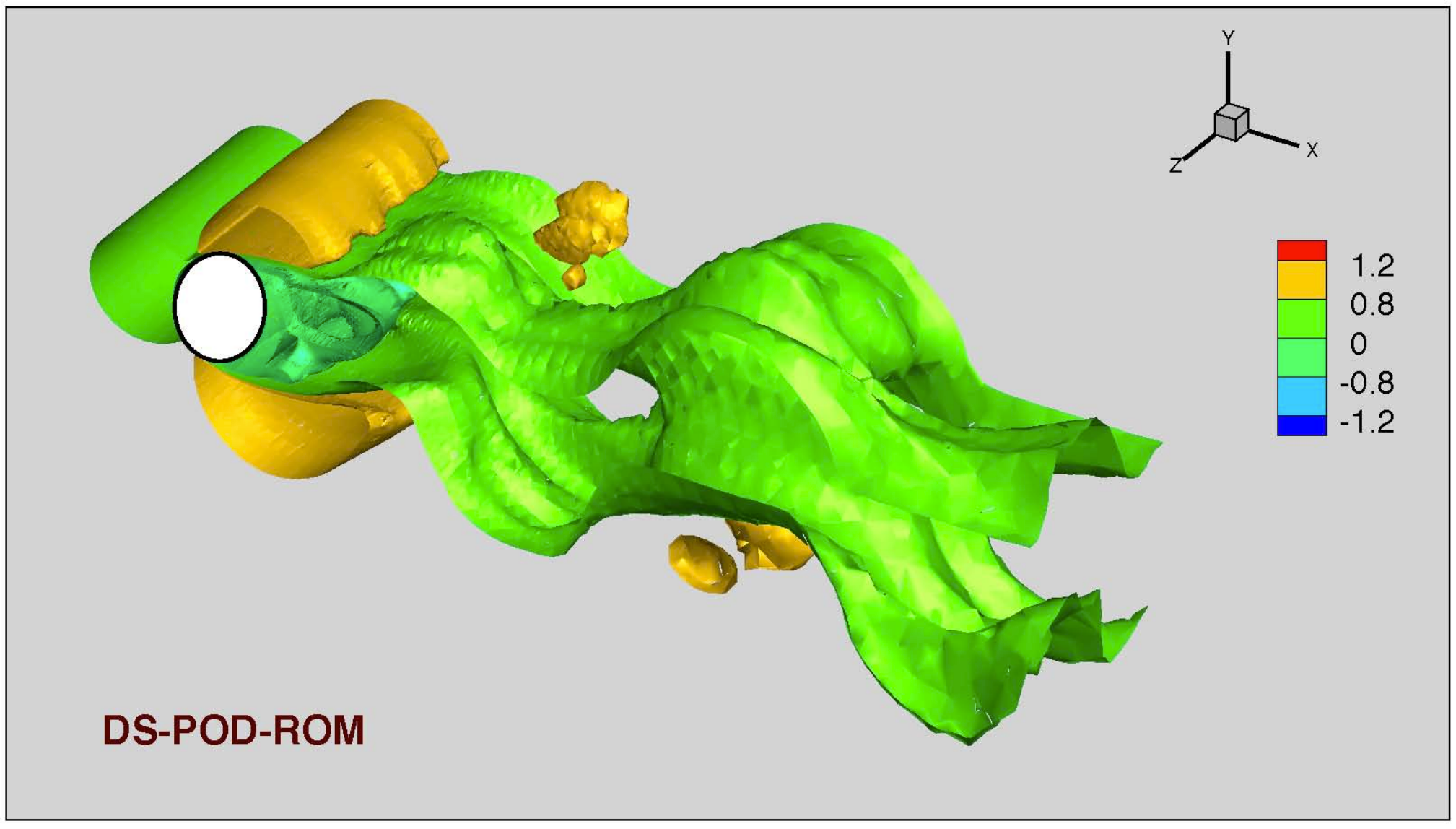}\end{minipage}
\end{figure}

\subsection{Numerical results}
\label{ss_numerical_results}

In this section, we test the four POD-ROMs described in \S\,\ref{s_closure_models}:
(i) the ML-POD-ROM \eqref{ML_POD_ROM_1}-\eqref{ML_POD_ROM_2}; 
(ii) the S-POD-ROM \eqref{S_POD_ROM_1}-\eqref{S_POD_ROM_2};
(iii) the new VMS-POD-ROM \eqref{VMS_POD_ROM_1}-\eqref{VMS_POD_ROM_9}; and
(iv) the new DS-POD-ROM \eqref{DS_POD_ROM_1}-\eqref{DS_POD_ROM_2}.
To assess their performance, we compare these four POD-ROMs with the 
POD-G-ROM \eqref{pod_g} (i.e., POD-ROM without any closure model) and 
the DNS projection of the evolution of the POD modes.
A successful POD-ROM should perform significantly better than the POD-G-ROM and
yield results that are close to those from the DNS.
The POD-ROM numerical results are judged according to the following two criteria: 
(i) the kinetic energy spectrum,  which represents the temporal average behavior of the POD-ROMs; and
(ii) the time evolution of the POD coefficients, which measures the instantaneous behavior of the POD-ROMs. 
We also include a computational efficiency assessment for all four POD-ROMs.

Before starting the quantitative comparison of the four POD-ROMs, we first give a flavor 
of the topology of the resulting flow fields.
Figure~\ref{fig_3d_snapshots} presents snapshots of horizontal velocity at $t=142.4\,s$ for
DNS, POD-G-ROM, ML-POD-ROM, S-POD-ROM, VMS-POD-ROM, and DS-POD-ROM.
For clarity, only five isosurfaces are drawn. 
Taking the DNS results as a benchmark, the POD-G-ROM seems to add unphysical structures.
The ML-POD-ROM, on the other hand, appears to add too much numerical dissipation to the
system and thus eliminates some of the vortical structures in the wake.
The S-POD-ROM, VMS-POD-ROM, and DS-POD-ROM perform well, capturing a similar amount
of structure as the DNS.
It also seems that there is some phase shift for all these POD-ROMs.
Due to space limitations, only one time instance snapshot is shown for the POD-ROMs.
The general behavior over the entire time interval is similar; it can be found at \url{http://www.math.vt.edu/people/wangzhu/POD_3DNumComp.html}.

\begin{figure}
\centering
\caption{
Kinetic energy spectrum of the DNS (blue) and the POD-ROMs (red):
(a) the POD-G-ROM \eqref{pod_g} overestimates the DNS spectrum;
(b) the ML-POD-ROM \eqref{ML_POD_ROM_1}-\eqref{ML_POD_ROM_2} underestimates
the DNS spectrum; 
(c) the S-POD-ROM \eqref{S_POD_ROM_1}-\eqref{S_POD_ROM_2} yields a more accurate
spectrum than the ML-POD-ROM, but displays high oscillations at the higher frequencies;
(d) the new VMS-POD-ROM \eqref{VMS_POD_ROM_1}-\eqref{VMS_POD_ROM_9} clearly
improves the accuracy of the S-POD-ROM's spectrum, displaying smaller oscillations at the
higher frequencies; and
(e) the new DS-POD-ROM \eqref{DS_POD_ROM_1}-\eqref{DS_POD_ROM_2} decreases 
the amplitude of the high oscillations of the VMS-POD-ROM's spectrum, although it displays
sporadic undershoots.
}
\label{fig_3d_spectrum}
\begin{minipage}[h]{0.03\linewidth} (a) \end{minipage}
\begin{minipage}[h]{0.7\linewidth} \includegraphics[width=1\textwidth]{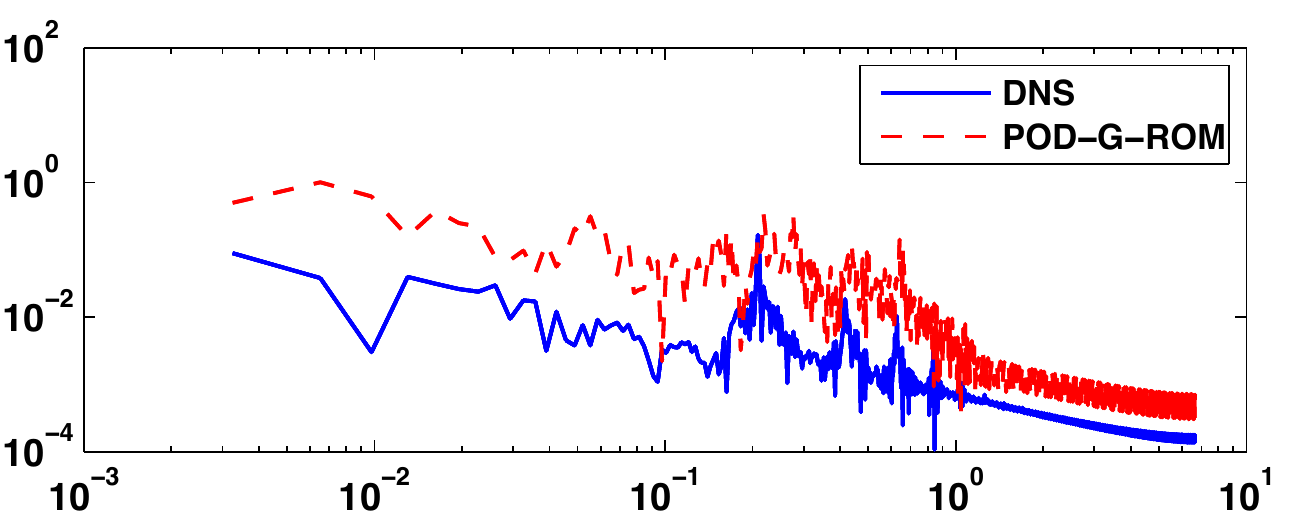}\end{minipage}\\
\begin{minipage}[h]{0.03\linewidth} (b) \end{minipage}
\begin{minipage}[h]{0.7\linewidth} \includegraphics[width=1\textwidth]{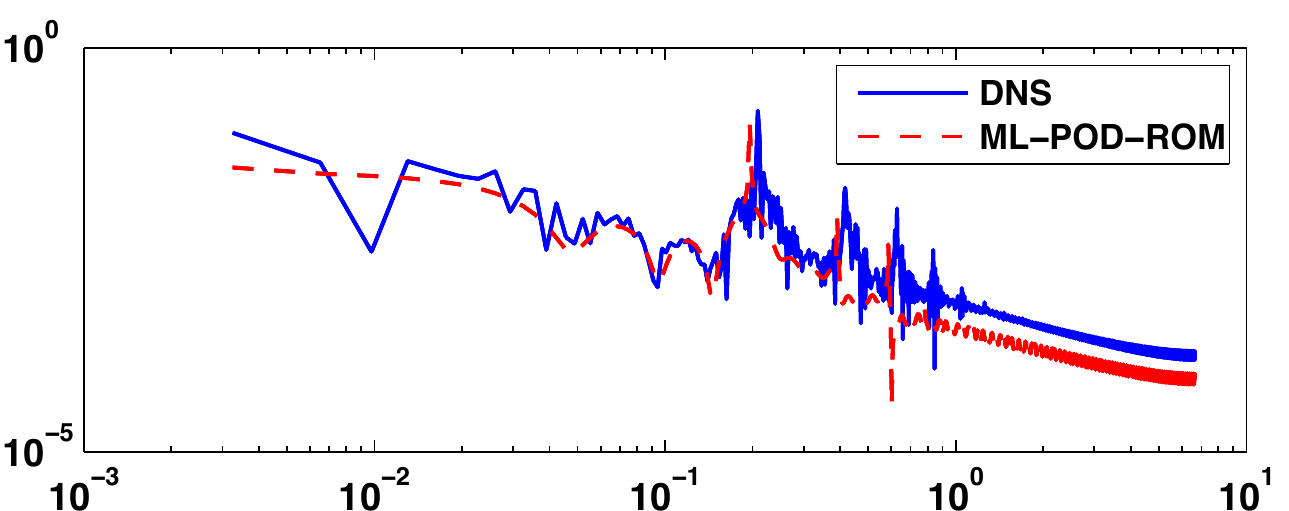}\end{minipage}\\
\begin{minipage}[h]{0.03\linewidth} (c) \end{minipage}
\begin{minipage}[h]{0.7\linewidth} \includegraphics[width=1\textwidth]{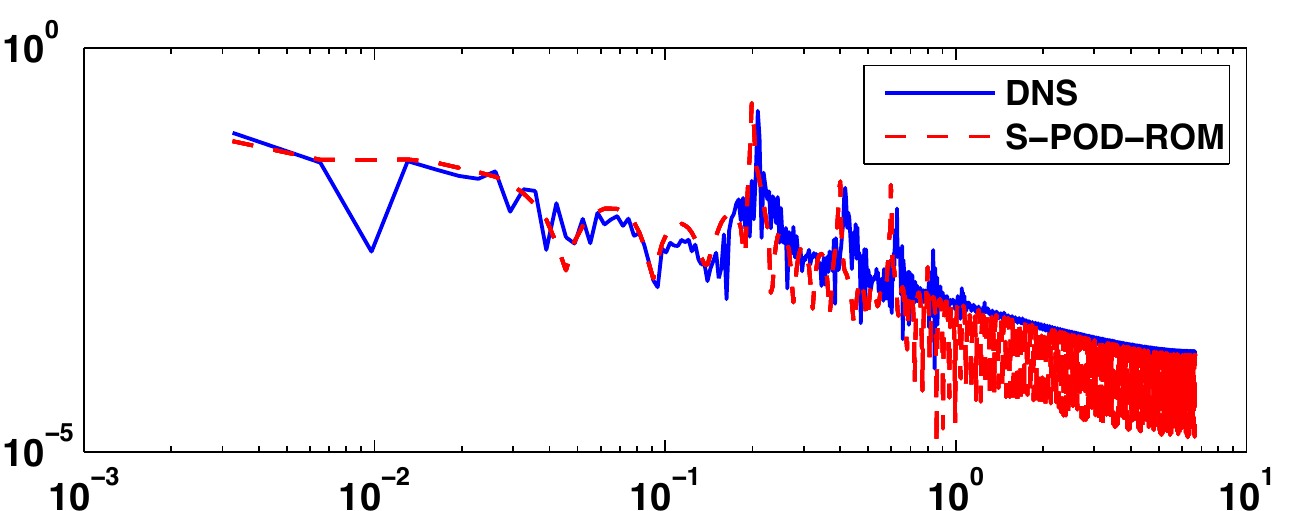}\end{minipage}\\
\begin{minipage}[h]{0.03\linewidth} (d) \end{minipage}
\begin{minipage}[h]{0.7\linewidth} \includegraphics[width=1\textwidth]{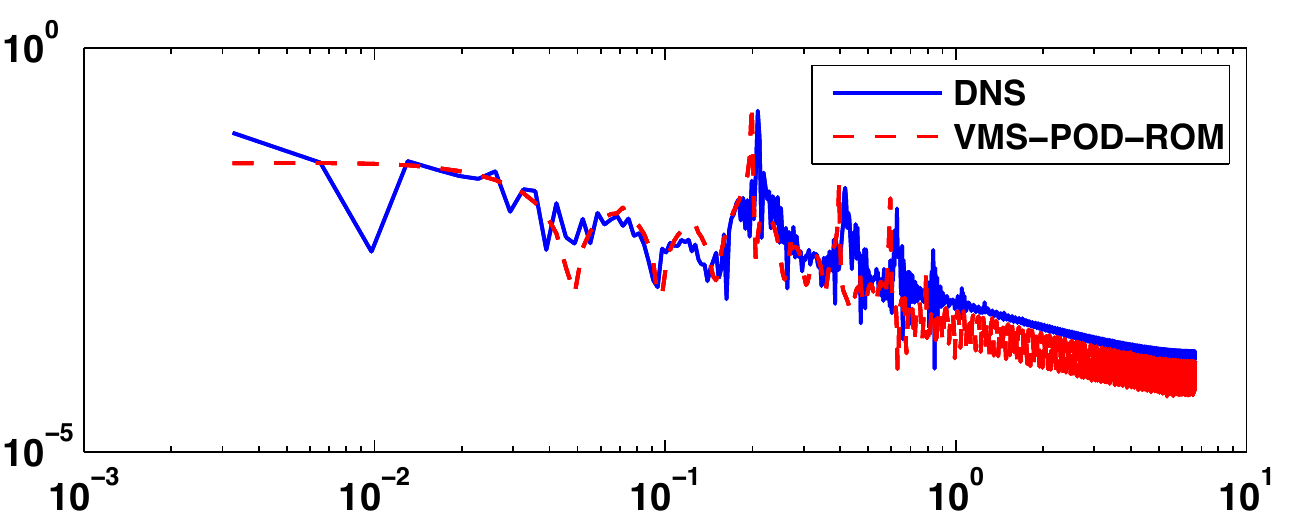}\end{minipage}\\
\begin{minipage}[h]{0.03\linewidth} (e) \end{minipage}
\begin{minipage}[h]{0.7\linewidth} \includegraphics[width=1\textwidth]{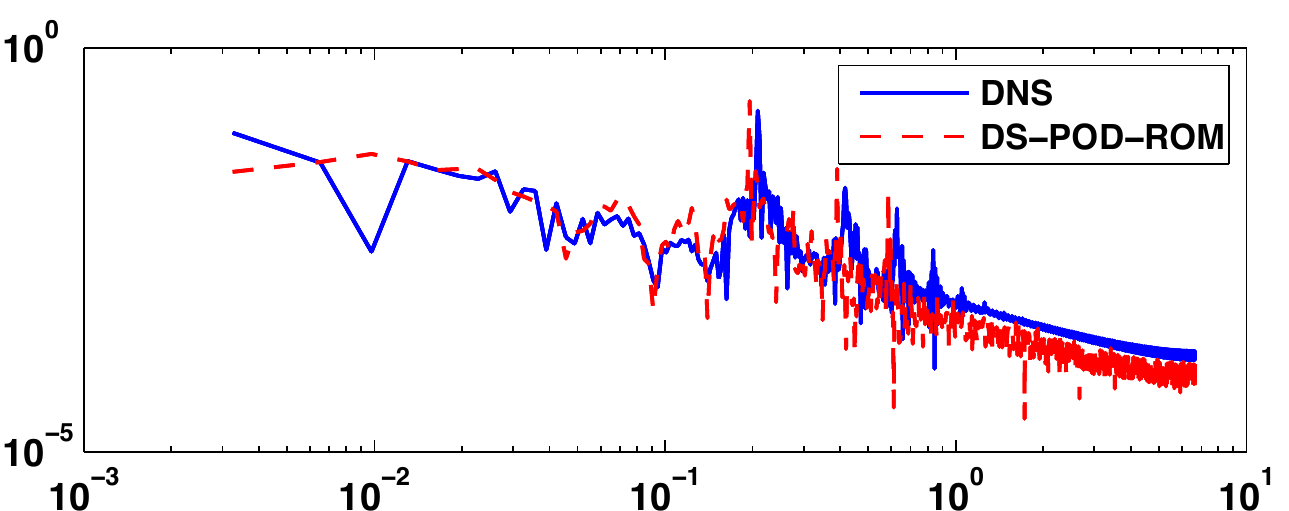}\end{minipage}
\end{figure}

Figure~\ref{fig_3d_spectrum} presents the energy spectra of the four POD-ROMs and, for comparison
purposes, of the POD-G-ROM.
The five energy spectra are compared with the DNS energy spectrum.
All energy spectra are calculated from the average kinetic energy of the nodes in the cube with side $0.1$ 
centered at the probe $(0.9992, 0.3575, 1.0625)$.
It is clear that the energy spectrum of the POD-G-ROM overestimates the energy spectrum of the DNS.
The energy spectrum of the ML-POD-ROM, on the other hand, underestimates the the energy spectrum 
of the DNS, especially at the higher frequencies.
The S-POD-ROM has a more accurate spectrum than the ML-POD-ROM, but it displays high oscillations
at the higher frequencies.
The VMS-POD-ROM is a clear improvement over the S-POD-ROM, with smaller oscillations at the higher 
frequencies.
The energy spectrum of the DS-POD-ROM is qualitatively similar to that of the VMS-POD-ROM. 
The DS-POD-ROM spectrum decreases the amplitude of the high frequency oscillations 
of the VMS-POD-ROM even further, although it introduces some sporadic large amplitude 
oscillations at high frequencies.
To summarize, the DS-POD-ROM and the VMS-POD-ROM yield the most accurate energy spectra, i.e.,
the closest to the DNS energy spectrum.
On the average, the DS-POD-ROM performs slightly better than the VMS-POD-ROM.

\begin{figure}
\centering
\caption{
Time evolutions of the POD basis coefficient $a_1$ of the DNS (blue) and the 
POD-ROMs (red):
(a) the POD-G-ROM \eqref{pod_g} overestimates the DNS results;
(b) the ML-POD-ROM \eqref{ML_POD_ROM_1}-\eqref{ML_POD_ROM_2} underestimates
the DNS results; 
(c) the S-POD-ROM \eqref{S_POD_ROM_1}-\eqref{S_POD_ROM_2} yields more accurate
results than the ML-POD-ROM;
(d) the new VMS-POD-ROM \eqref{VMS_POD_ROM_1}-\eqref{VMS_POD_ROM_9} 
improves the accuracy of the S-POD-ROM results, especially toward the end of the
simulation; and
(e) the new DS-POD-ROM \eqref{DS_POD_ROM_1}-\eqref{DS_POD_ROM_2} yields
results that are similar to those of the VMS-POD-ROM.
}
\label{fig_3d_evo_a1}
\begin{minipage}[h]{0.03\linewidth} (a) \end{minipage}
\begin{minipage}[h]{0.7\linewidth} \includegraphics[width=1\textwidth]{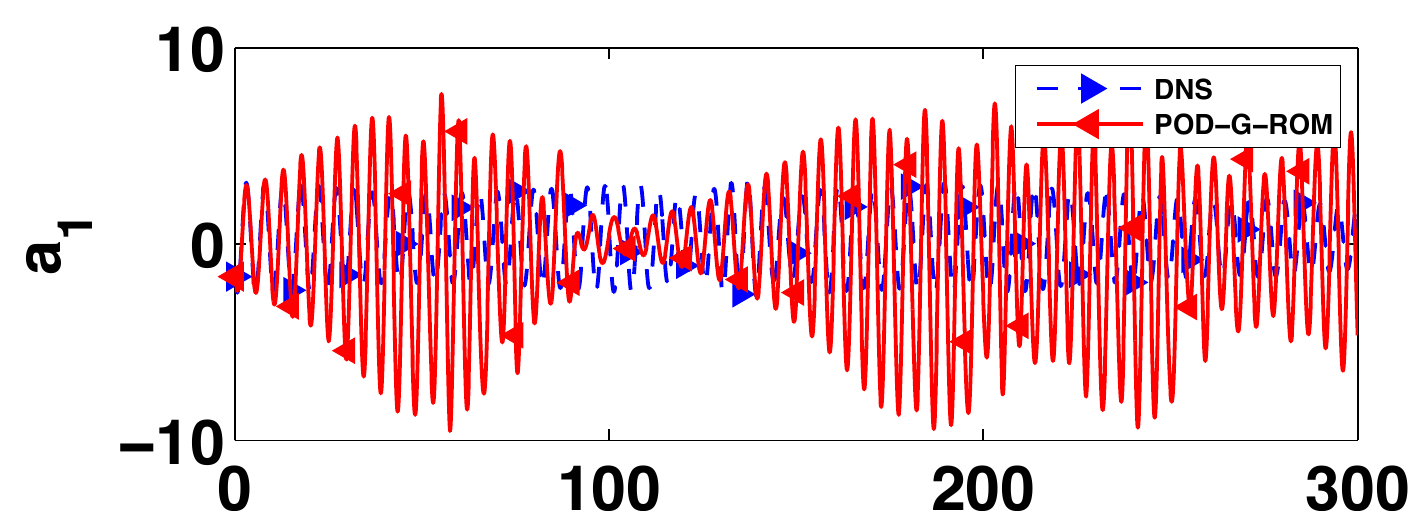} \end{minipage}\\
\begin{minipage}[h]{0.03\linewidth} (b) \end{minipage}
\begin{minipage}[h]{0.7\linewidth} \includegraphics[width=1\textwidth]{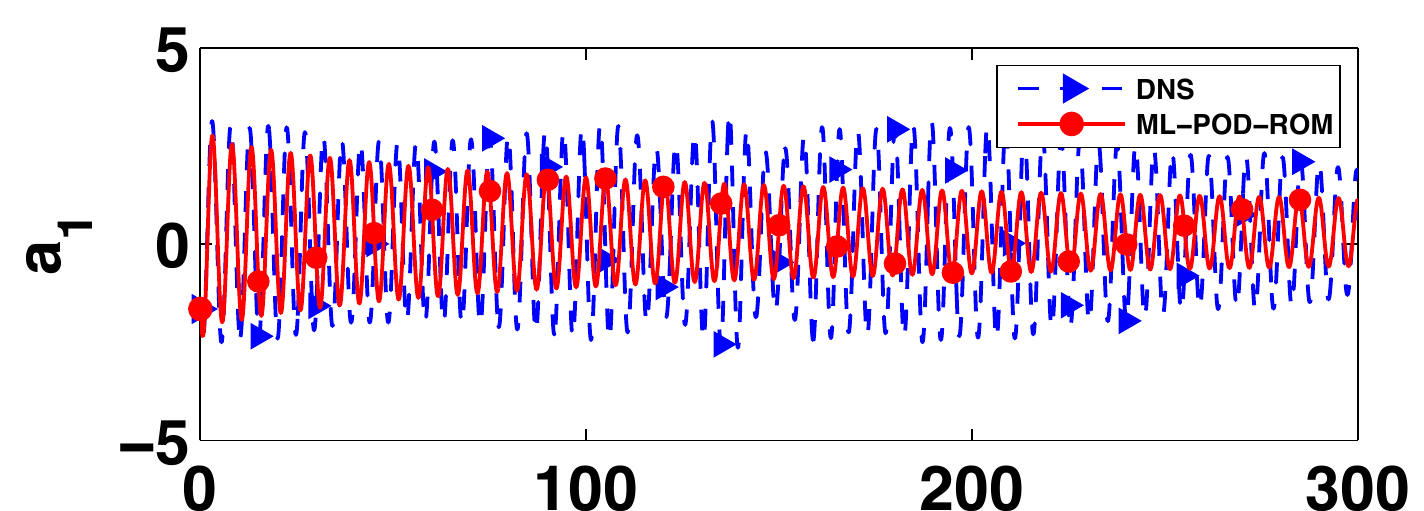} \end{minipage}\\
\begin{minipage}[h]{0.03\linewidth} (c) \end{minipage}
\begin{minipage}[h]{0.7\linewidth} \includegraphics[width=1\textwidth]{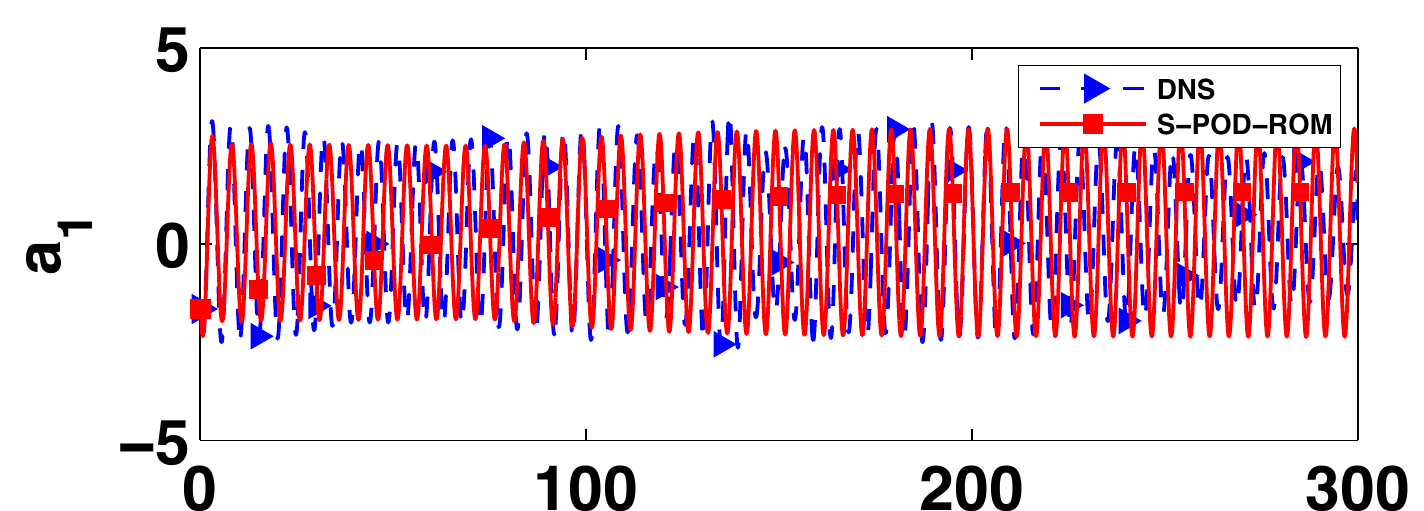} \end{minipage}\\
\begin{minipage}[h]{0.03\linewidth} (d) \end{minipage}
\begin{minipage}[h]{0.7\linewidth} \includegraphics[width=1\textwidth]{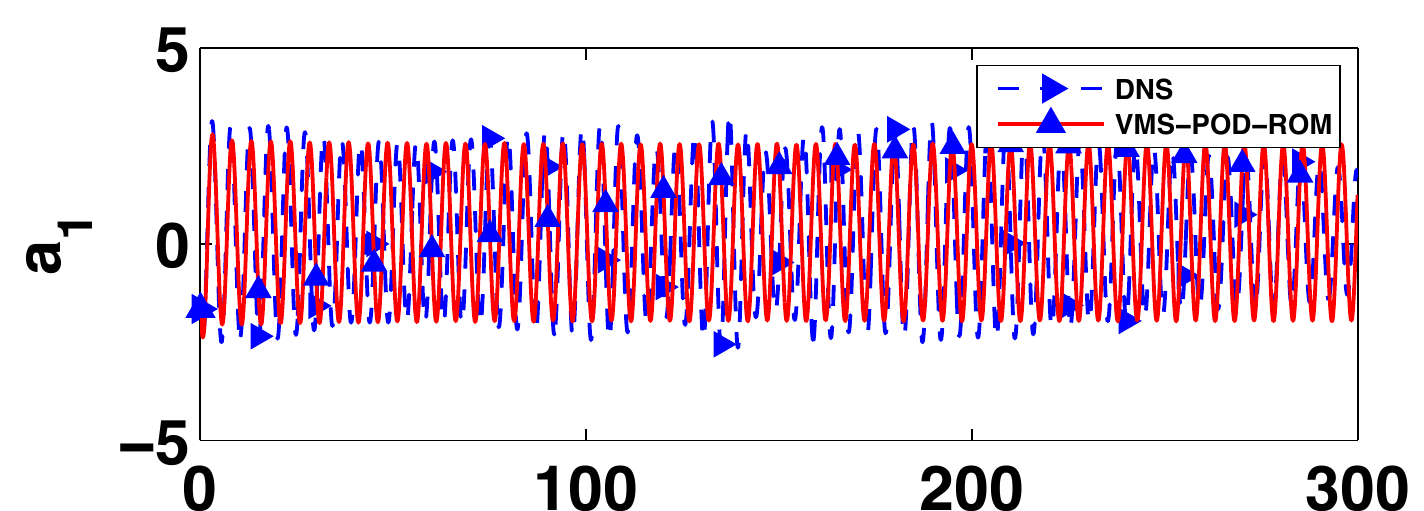} \end{minipage}\\
\begin{minipage}[h]{0.03\linewidth} (e) \end{minipage}
\begin{minipage}[h]{0.7\linewidth} \includegraphics[width=1\textwidth]{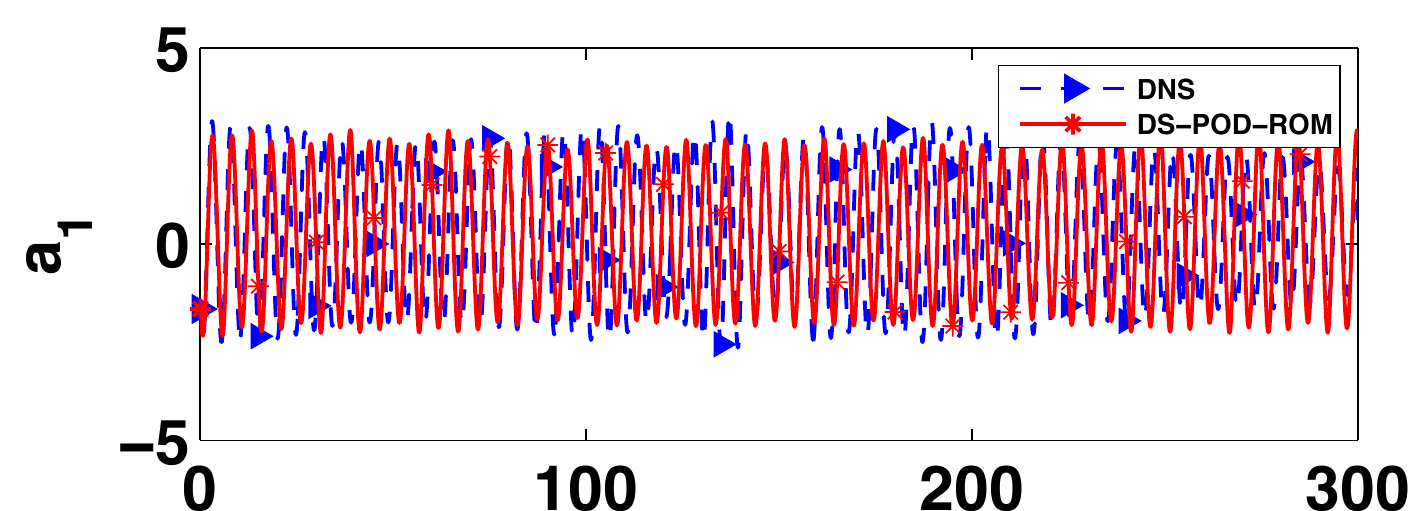} \end{minipage}\\
\end{figure}

\begin{figure}
\centering
\caption{
Time evolutions of the POD basis coefficient $a_4$ of the DNS (blue) and the 
POD-ROMs (red):
(a) the POD-G-ROM \eqref{pod_g} overestimates the DNS results;
(b) the ML-POD-ROM \eqref{ML_POD_ROM_1}-\eqref{ML_POD_ROM_2} underestimates
the DNS results; 
(c) the S-POD-ROM \eqref{S_POD_ROM_1}-\eqref{S_POD_ROM_2} yields more accurate
results than the ML-POD-ROM;
(d) the new VMS-POD-ROM \eqref{VMS_POD_ROM_1}-\eqref{VMS_POD_ROM_9} 
improves the accuracy of the S-POD-ROM results, especially toward the end of the
simulation; and
(e) the new DS-POD-ROM \eqref{DS_POD_ROM_1}-\eqref{DS_POD_ROM_2} performs
slightly better than the VMS-POD-ROM, recovering some of the variability of the DNS results.
}
\label{fig_3d_evo_a4}
\begin{minipage}[h]{0.03\linewidth} (a) \end{minipage}
\begin{minipage}[h]{0.7\linewidth} \includegraphics[width=1\textwidth]{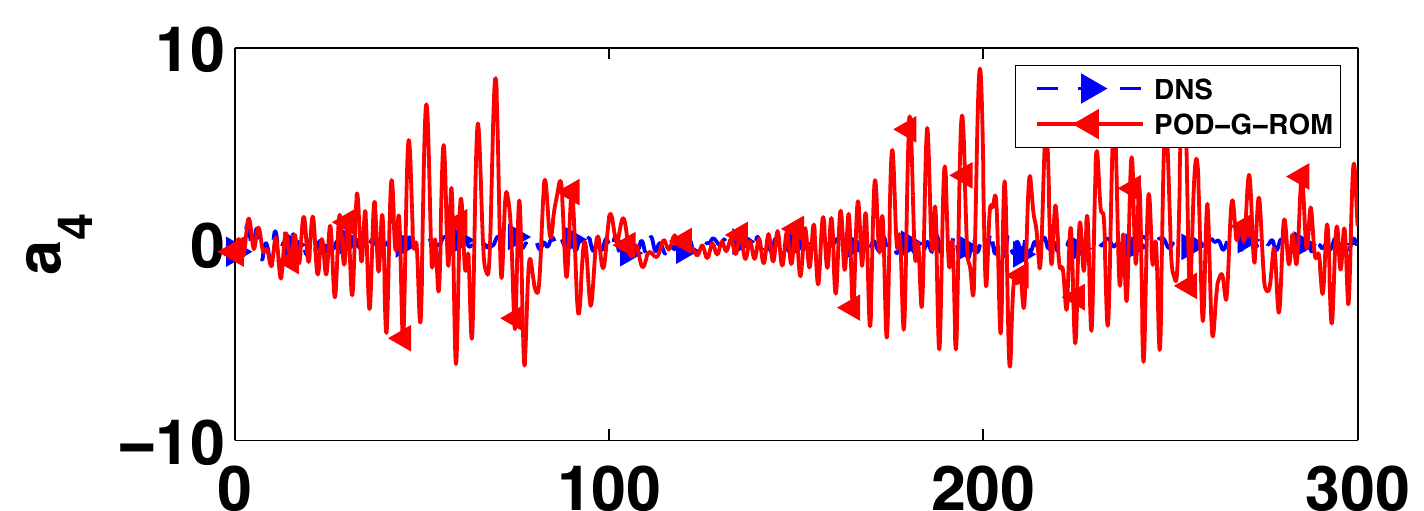} \end{minipage}\\
\begin{minipage}[h]{0.03\linewidth} (b) \end{minipage}
\begin{minipage}[h]{0.7\linewidth} \includegraphics[width=1\textwidth]{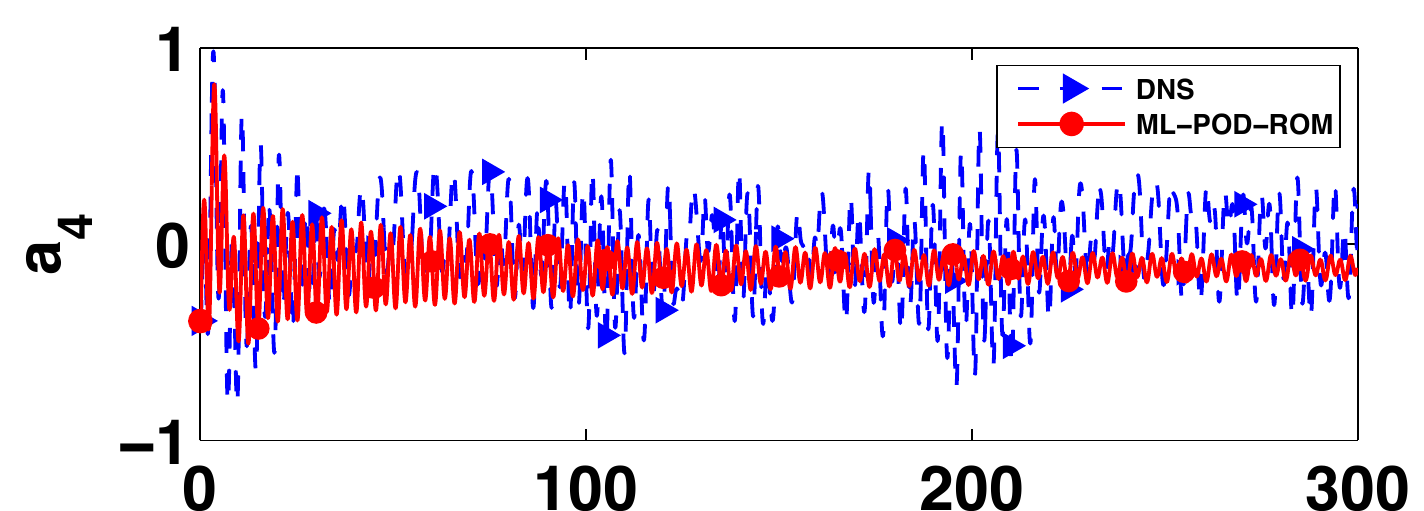} \end{minipage}\\
\begin{minipage}[h]{0.03\linewidth} (c) \end{minipage}
\begin{minipage}[h]{0.7\linewidth} \includegraphics[width=1\textwidth]{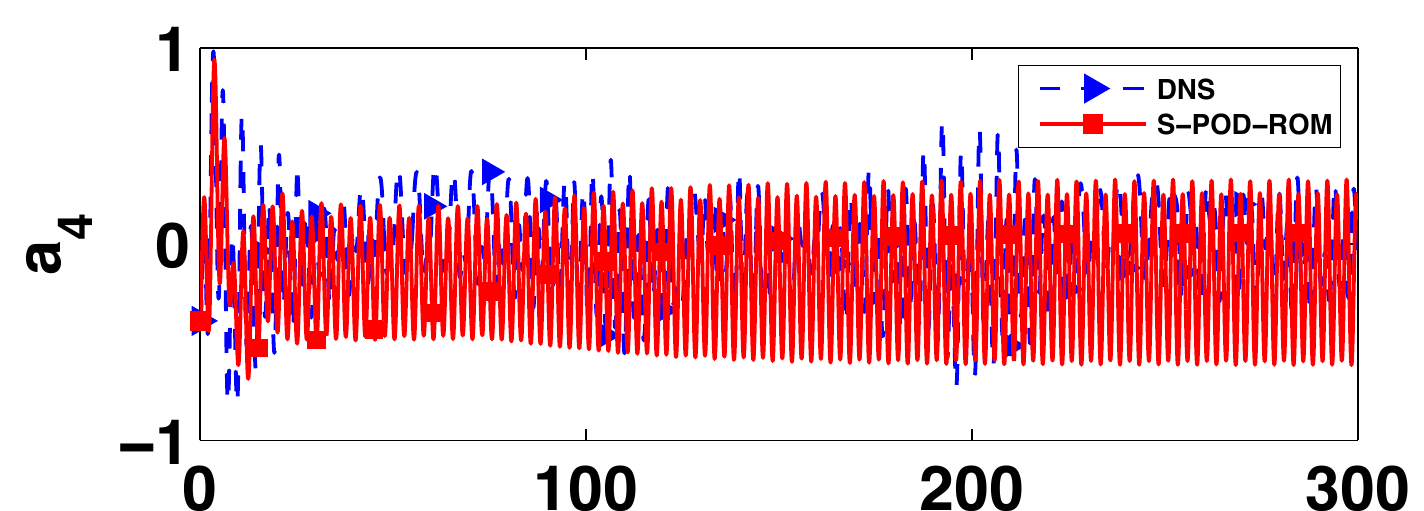} \end{minipage}\\
\begin{minipage}[h]{0.03\linewidth} (d) \end{minipage}
\begin{minipage}[h]{0.7\linewidth} \includegraphics[width=1\textwidth]{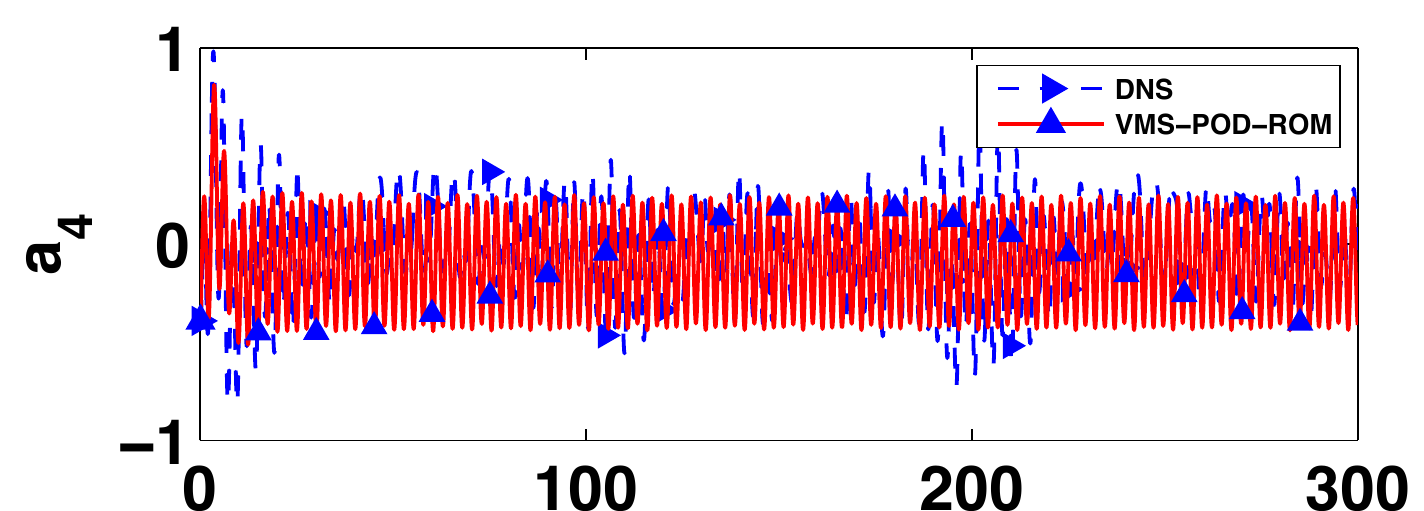} \end{minipage}\\
\begin{minipage}[h]{0.03\linewidth} (e) \end{minipage}
\begin{minipage}[h]{0.7\linewidth} \includegraphics[width=1\textwidth]{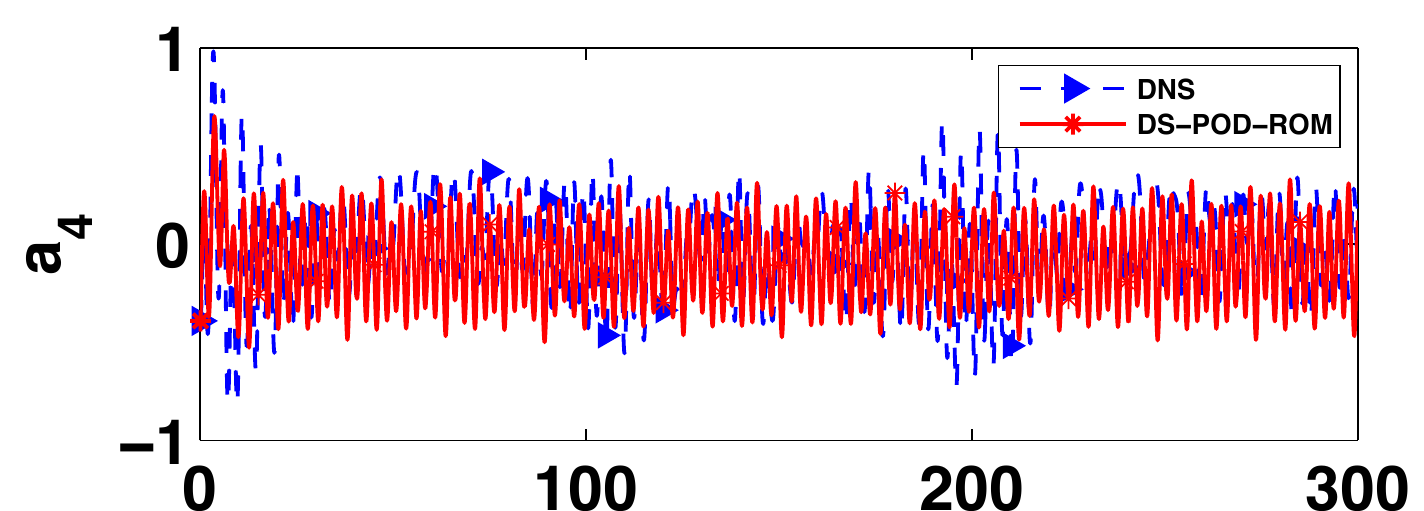} \end{minipage}
\end{figure}

As the second criterion in judging the performance of the four POD-ROMs, the time evolutions of the POD 
basis coefficients $a_1(\cdot)$ and $a_4(\cdot)$ on the entire time interval $[0, 300]$ are shown in 
Figures~\ref{fig_3d_evo_a1}-\ref{fig_3d_evo_a4}. 
We note that the other POD coefficients have a similar behavior. 
Thus, for clarity of exposition, we include only 
$a_1(\cdot)$ and $a_4(\cdot)$. 
The POD-G-ROM's time evolutions of $a_1$ and $a_4$ are clearly inaccurate. 
Indeed, the magnitude of $a_4$ is nine times larger than that of the DNS projection, which indicates the 
need for closure modeling. 
The ML-POD-ROM's time evolutions of $a_1$ and $a_4$ are also inaccurate. 
Specifically, although the time evolution at the beginning of the simulation (where the EV constant $\alpha$ 
was chosen) is relatively accurate, the accuracy significantly degrades toward the end of the simulation.
For example, the magnitude of $a_4$ at the end of the simulation is only one eighth of that of the DNS.  
The S-POD-ROM yields more accurate time evolutions than the ML-POD-ROM for both $a_1$ and $a_4$,
although the magnitude of the POD coefficients stays almost constant at a high level.
The VMS-POD-ROM's time evolutions of $a_1$ and $a_4$ are better than those of the S-POD-ROM,
since the magnitudes of the POD coefficients are closer to those of the DNS.
Finally, the DS-POD-ROM also yields accurate results.
We note that the DS-POD-ROM's $a_1$ and $a_4$ coefficients have significantly more variability than
the corresponding coefficients of the VMS-POD-ROM.
This is a consequence of the fact that the EV coefficient $C_S$ varies in time and space for the DS-POD-ROM,
whereas it is constant for the VMS-POD-ROM.
To summarize, the DS-POD-ROM and the VMS-POD-ROM perform the best.
On the average, the DS-POD-ROM performs slightly better than the VMS-POD-ROM.

Based on the energy spectra and the the time evolutions of the POD basis coefficients 
$a_1(\cdot)$ and $a_4(\cdot)$, the DS-POD-ROM and the VMS-POD-ROM consistently 
outperform the ML-POD-ROM and the S-POD-ROM.
To determine which one of the DS-POD-ROM and the VMS-POD-ROM performs best, we
collected the results in Figures~\ref{fig_3d_evo_a4}(d)--\ref{fig_3d_evo_a4}(e) 
(corresponding to the time evolution of the POD basis coefficient $a_4(\cdot)$ for the
DNS projection, the VMS-POD-ROM and the DS-POD-ROM) and we displayed them in the 
same plot in Figure~\ref{fig_3d_comp}.
Since it is difficult to distinguish between the results from the VMS-POD-ROM and the DS-POD-ROM,
we zoomed in on the POD basis coefficient $a_4$ over the time interval $[266, 282]$. 
Based on the plot in the inset, it is clear that, for this time interval, the DS-POD-ROM performs better 
than the VMS-POD-ROM.
More importantly, it appears that the magnitude of $a_4$ in the DS-POD-ROM displays some
of the variability displayed by the DNS;  the magnitude of the VMS-POD-ROM's $a_4$ coefficient, 
on the other hand, displays an almost periodic behavior.
We believe that the variation of the DS-POD-ROM's $a_4$ coefficient is due to the dynamical
computation of the EV coefficient, which changes both in space and time; the EV coefficient of
the VMS-POD-ROM, however, is constant and is computed at the beginning of the simulation.

To gain further insight into the behavior of the DS-POD-ROM and the VMS-POD-ROM, we 
considered the root mean square horizontal velocity of the two models.
Figure~\ref{fig_3d_avg} presents the average horizontal velocity $\left<u\right>$, 
which is computed at points with coordinates $x=3.2937$ and $y= 2.2796$, and the root mean 
square horizontal velocity $u_{rms}=\left<u-\left<u\right>, u-\left<u\right>\right> ^{1/2} / \left<u\right>$, 
where $\left<\cdot\right>$ here denotes the spatial average in the $z$-direction.
Both the DS-POD-ROM and the VMS-POD-ROM yield average horizontal and root mean square 
velocities that are in close agreement with the DNS data.
As for the time evolution of the POD basis coefficient $a_4$ in Figure~\ref{fig_3d_comp}, the 
DS-POD-ROM and the VMS-POD-ROM results are practically indistinguishable.

To summarize, the VMS and DS approaches, which are state-of-the-art closure modeling 
strategies in LES, yield the most accurate POD closure models for the 3D turbulent 
flow that we investigated.
A natural question, however, is whether the new POD closure modeling strategies that we
proposed display a high level of computational efficiency, which is one of the trademarks 
of a successful POD-ROM.
To answer this question, we computed the CPU times for all four POD-ROMs and compared
them with those of the DNS and the POD-G-ROM.

To make such a comparison, however, we first need to address the numerical differences between 
the DNS and the POD-ROMs. 
First, the discretizations used in the two approaches are completely different. 
Indeed, the spatial discretization used in the DNS was the finite volume method, whereas for the 
POD-ROMs we used a finite element method. 
Furthermore, the time-discretization used in the DNS was second-order (Crank-Nicolson and 
Adams-Bashforth), whereas in the POD-ROMs we used a first-order time discretization (explicit Euler). 
The time steps employed were also different: $\Delta t=2\times 10^{-3}$ in the DNS and 
$\Delta t=7.5\times 10^{-4}$ in the POD-ROM. 
Most importantly, the DNS was performed on a parallel machine (on $16$ processors), whereas all the 
POD-ROM runs were carried out on a single-processor machine. 
Thus, to ensure a more realistic comparison between the CPU times of the DNS and the POD-ROMs, we multiplied
the CPU time of the DNS by a factor of $16$.

\begin{table}
  \begin{center}
  		\caption{Speed-up factors of POD-ROMs. }
		\label{speed-up-factor}
	\begin{tabular}{cccccc}
		\hline
		{}&\,POD-G-ROM\,&\,ML-POD-ROM\,&\,S-POD-ROM\,&\,VMS-POD-ROM\,&\,DS-POD-ROM \,\\
		\hline
		$S_f$&665&659&36&41&23\\
		\hline
	\end{tabular}
  \end{center}
\end{table}

To measure the computational efficiency of the four POD-ROMs, we define the 
speed-up factor
\begin{equation}
S_f=\frac{\text{CPU time of DNS}}{\text{CPU time of POD-ROM}}
\end{equation}
and list results in Table \ref{speed-up-factor}.
The most efficient model is the POD-G-ROM.
This is not surprising, since no closure model is used in POD-G-ROM and 
thus no CPU time is spent computing an additional nonlinear term at each time step.
The second most efficient model is the ML-POD-ROM.
This is again natural, since only a linear closure model is employed in the ML-POD-ROM
and thus the computational overhead is minimal.
The speed-up factors for the S-POD-ROM, the VMS-POD-ROM and the DS-POD-ROM are one
order of magnitude lower than those for the ML-POD-ROM and the POD-G-ROM.
The reason is that the former use nonlinear closure models, which increase significantly the 
computational time.
Note, however, that the S-POD-ROM, the VMS-POD-ROM and the DS-POD-ROM are still 
significantly more efficient than the DNS.
\begin{landscape}
\begin{figure}
\centering
\caption{
Time evolutions of the POD basis coefficient $a_4$ of the DNS (green), 
the new VMS-POD-ROM \eqref{VMS_POD_ROM_1}-\eqref{VMS_POD_ROM_9} (blue), and
the new DS-POD-ROM \eqref{DS_POD_ROM_1}-\eqref{DS_POD_ROM_2} (red).
The inset shows that the DS-POD-ROM performs better than the VMS-POD-ROM, capturing
some of the variability displayed by the DNS results.
}
\label{fig_3d_comp}
\begin{minipage}[ht]{1\linewidth}
\hspace{-1cm}
\includegraphics[width=1.0\textwidth]{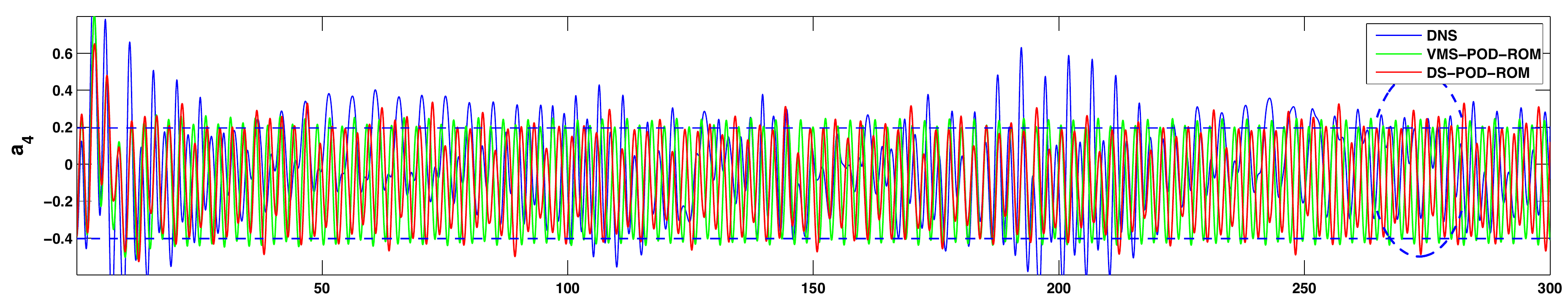}
\end{minipage}
\setlength{\unitlength}{1mm}
{\color{blue}
\begin{picture}(10, 10)
  \thicklines
    \put(70, 10){\vector(-1, -1){15}}
\end{picture}
}
\begin{minipage}[ht]{1\linewidth}
\hspace{5cm}
\includegraphics[width=0.35\textwidth]{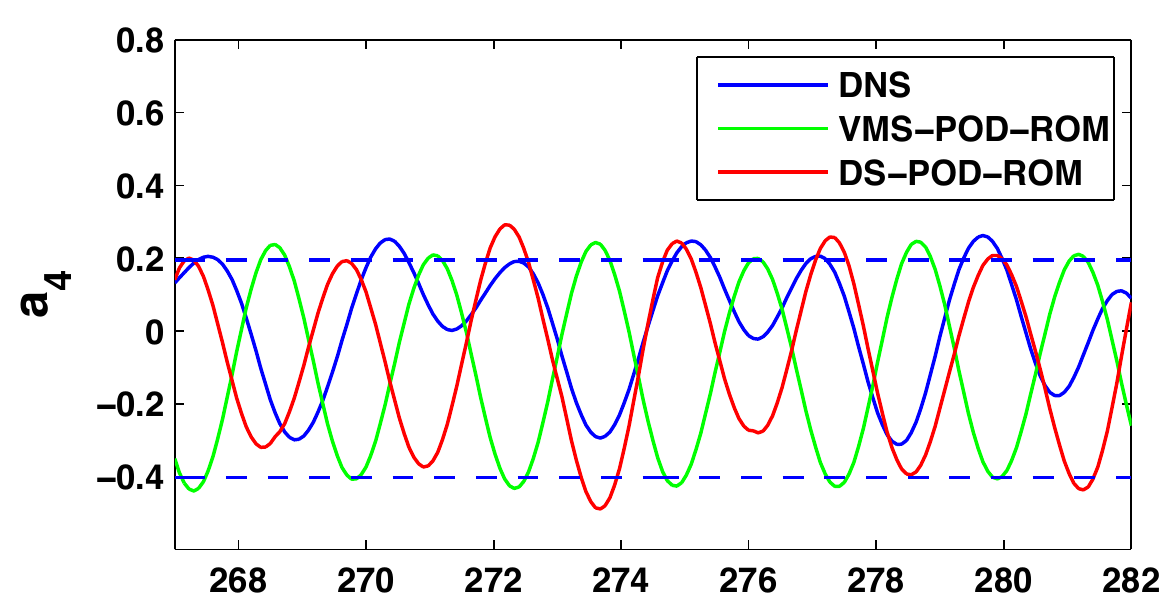}
\end{minipage}
\end{figure}
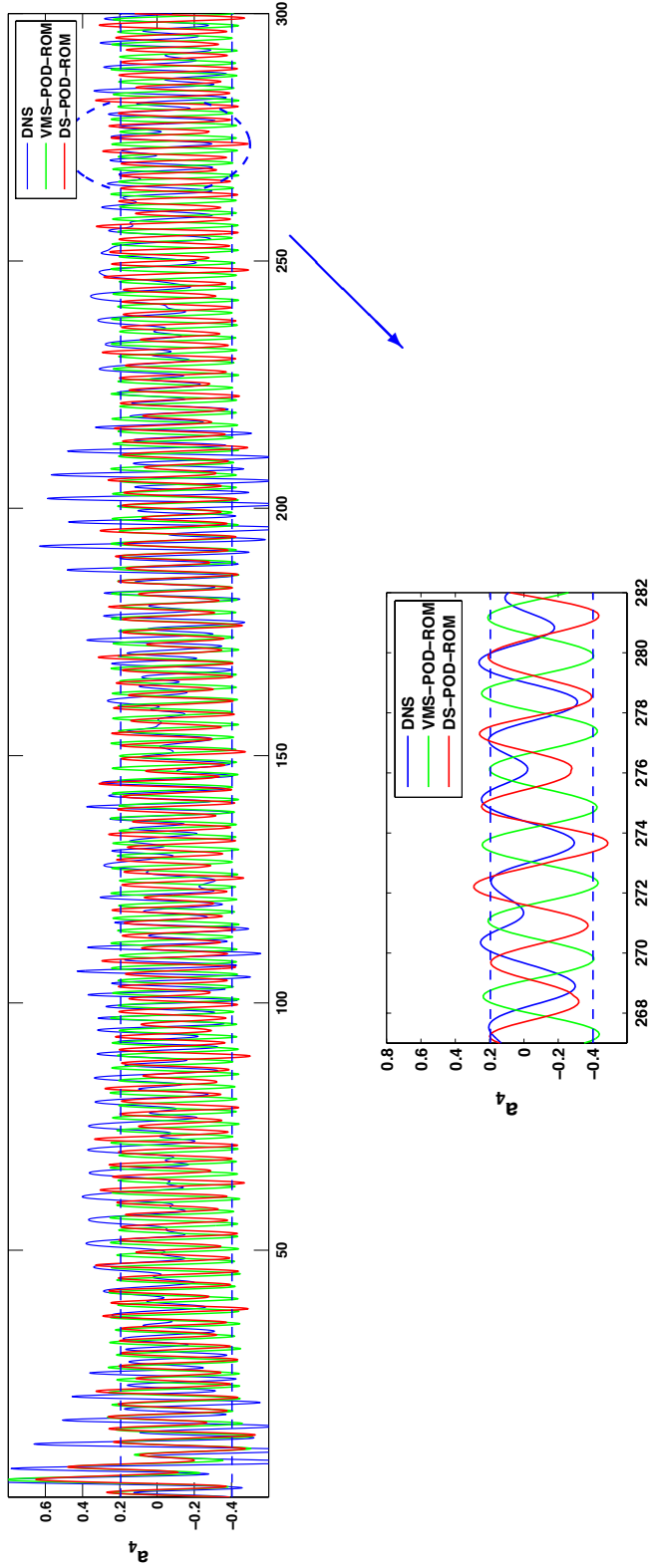
%
\begin{figure}
\centering
\caption{
Time evolutions of the average horizontal velocity (top) and 
the root mean square horizontal velocity (bottom)
of the DNS (green), 
the new VMS-POD-ROM \eqref{VMS_POD_ROM_1}-\eqref{VMS_POD_ROM_9} (blue), and
the new DS-POD-ROM \eqref{DS_POD_ROM_1}-\eqref{DS_POD_ROM_2} (red).
The DS-POD-ROM and the VMS-POD-ROM yield similar results.
}
\label{fig_3d_avg}
\begin{minipage}[ht]{1\linewidth}
\hspace{-1cm}
\includegraphics[width=1.0\textwidth]{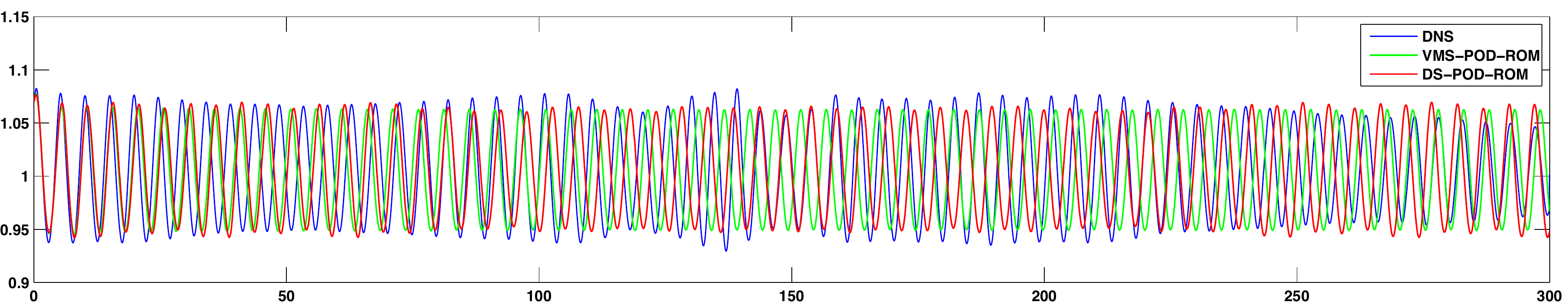}
\end{minipage}
\\\vspace{1cm}
\begin{minipage}[ht]{1\linewidth}
\hspace{-1cm}
\includegraphics[width=1.0\textwidth]{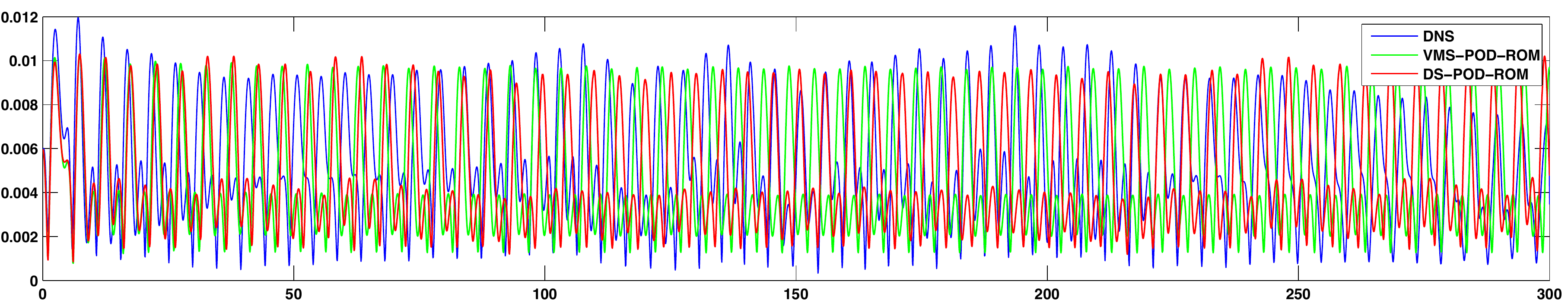}
\end{minipage}
\end{figure}
\end{landscape}

\section{Conclusions}
\label{s_conclusions}

This paper put forth two new POD-ROMs (the DS-POD-ROM and the VMS-POD-ROM), 
which are inspired from state-of-the-art LES closure modeling strategies.
These two new POD-ROMs together with the ML-POD-ROM and the S-POD-ROM were 
tested in the numerical simulation of a 3D turbulent flow past a cylinder at $Re =1,000$.
For completeness, we also included results with the POD-G-ROM (i.e., a POD-ROM without 
any closure model), as well as the DNS projection of the evolution of the POD modes, which 
served as benchmark for our numerical simulations. 

To assess the performance of the POD-ROMs, two criteria were considered in this paper: 
the kinetic energy spectrum and the time evolution of the POD basis coefficients. 
The former is used to measure the average behavior of the POD-ROMs and the latter is used 
to quantify the instantaneous behavior of these models. 
Both the POD-G-ROM and the ML-POD-ROM yielded inaccurate results. 
The DS-POD-ROM and the VMS-POD-ROM clearly outperformed these two models, yielding
more accurate results for both the kinetic energy spectrum and the time evolution of the POD basis 
coefficients. 
The DS-POD-ROM performed slightly better than the VMS-POD-ROM for both criteria and also
seemed to display more adaptivity in terms of adjusting the magnitude of the POD basis coefficients.
Overall, however, the two models yielded similar qualitative results.
This seems to reflect the LES setting, where both the DS and the VMS closure modeling strategies
are considered state-of-the-art \cite[]{HMOW01,HOM01}.
The DS-POD-ROM and the VMS-POD-ROM, although not as computationally efficient as the 
POD-G-ROM or the ML-POD-ROM, significantly decreased the CPU time of the DNS.
To summarize, for the 3D turbulent flow that we investigated, the DS-POD-ROM and the 
VMS-POD-ROM were found to perform the best among the POD-ROMs investigated, combining 
a relative high numerical accuracy with a high level of computational efficiency.

%
 
We plan to further investigate several other research avenues. 
First, we plan to study more efficient time-discretization approaches and take advantage of 
parallel computing in order to further decrease the computational time and, at the same time, 
increase the dimension (and the thus physical accuracy) of the POD-ROMs. 
Second, since the linear closure model (ML-POD-ROM) is computationally efficient, but only 
works on a relative short time interval if the appropriate EV coefficient $\alpha$ is chosen, 
we will investigate a hybrid approach: 
We will use the DS approach to calculate $\alpha$ only when the flow displays a high level of
variability, and then use this value in the ML-POD-ROM as long as the flow does not experience
sudden transitions. 
Third, using these computational developments, we will investigate the new POD-ROMs
in more challenging, higher Reynolds number structurally dominated turbulent flows.
Finally, we plan to employ the new POD-ROMs in other scientific and engineering applications 
in which accurate POD closure modeling is needed, such as optimal control, optimization, 
and data assimilation problems.

\begin{acknowledgments}
We greatly appreciate the financial support of the Air Force Office 
of Scientific Research through grant FA9550-08-1-0136 
and of the National Science Foundation through grant DMS-1016450.
A significant part of the computations were carried out on SystemX 
at Virginia Tech's Advanced Research Computing center 
(\url{http://www.arc.vt.edu}). 
The allocation grant and support provided by the staff are gratefully 
acknowledged.
\end{acknowledgments}

\newpage

\bibliographystyle{jfm2}
\bibliography{comprehensive_bibliography}

\end{document}